# Highlights

## Direct numerical simulations of turbulent drag reduction via piezoelectric actuation

Amir Amjadimanesh, Aman Kidanemariam, David Chappell, Mahdi Bodaghi, Amirreza Rouhi

- Realistic piezoelectric surface deformations are studied in a turbulent channel flow using DNS.
- Hybrid spanwise waves achieve up to 27.6% turbulent drag reduction.
- Transverse shear disrupts the near-wall turbulence regeneration cycle.

# Direct numerical simulations of turbulent drag reduction via piezoelectric actuation

Amir Amjadimanesh[a,*], Aman Kidanemariam[b], David Chappell[c], Mahdi Bodaghi[a] and Amirreza Rouhi[a]

[a]Department of Engineering, Nottingham Trent University, Clifton, Nottingham, NG11 8NS, Nottinghamshire, UK
[b]Department of Mechanical Engineering, University of Melbourne, Grattan Street, Melbourne, 3010, Victoria, Australia
[c]School of Science & Technology, Nottingham Trent University, Clifton, Nottingham, NG11 8NS, Nottinghamshire, UK

ABSTRACT

We have conducted Direct Numerical Simulations of turbulent half-channel flow over realistic surface deformations at friction Reynolds number $Re_\tau = 200$. We generated the surface deformations using piezoelectric actuators. We simulated the piezoelectric actuation over the practical actuation frequency range ($119Hz \leq f_{\rm act} \leq 543Hz$) and voltage range ($250V \leq Q \leq 500V$) beneath an Aluminum sheet using Finite Element Analysis. The sheet deformation amplitude and actuation frequency in viscous units vary within the range $2 \leq \eta^+_{\rm max} \leq 34$, and $-0.58 \leq \omega^+ \leq 0.70$. The vertical surface deformations from our actuation setup generate three types of waves: travelling, hybrid, and standing waves. Surface deformations are applied as bottom-wall boundary conditions of the turbulent channel flow to generate waves in the upstream, downstream, and spanwise directions. We achieved maximum drag reductions of 1.6%, 5.4%, and 27.6% for upstream, downstream, and spanwise waves, respectively. The streamwise waves generate alternating adverse and favorable pressure gradients, which locally increase and decrease drag, leading to a marginal net change in drag. In contrast, spanwise waves introduce transverse shear, accompanied by high- and low-streamwise-momentum zones that respectively attenuate and energize the near-wall turbulence. Such disruption of the near-wall turbulence-regeneration cycle produces up to 27% drag reduction for the realistic spanwise hybrid wave; such an outcome demonstrates the efficacy of unconventional realistic surface deformations in achieving significant drag reduction.

## 1. Intrduction

Drag is a primary source of energy loss in many industrial applications. For Class 8 heavy-duty vehicles on highways, aerodynamic drag accounts for approximately 65% of their total energy consumption (Sovran 1983; McCallen et al. 1999), with the remainder used to overcome rolling resistance and operate auxiliary systems. For a typical truck trailer, the ratio of skin-friction drag to total drag is 15%, and the remainder is due to form drag. In applications such as aircraft, submarines, and pipelines—where rolling resistance is absent—an even greater proportion of energy is lost to skin friction drag. For instance, 52% of the total drag is wasted due to skin friction drag in the Airbus A-380 (Smith et al. 2025). For other applications such as submarines and pipelines, the contribution of skin-friction drag increases to about 85% and 100%, respectively (Ricco et al. 2021; Renilson 2015). Therefore, even a marginal reduction in skin-friction drag (for brevity, we denote "turbulent skin-friction drag" as "drag") can directly yield significant energy savings. Hence, researchers have studied different techniques to reduce drag. Passive techniques, which are static non-deforming structures, such as riblets (Garcia-Mayoral and Jimenez 2011; García-Mayoral and Jiménez 2012; Endrikat et al. 2021,b; Modesti et al. 2021; Endrikat et al. 2022; Rouhi et al. 2022) and superhydrophobic surfaces (Min and Kim 2004; Rothstein 2010; Daniello et al. 2009; Lee and Kim 2011; Park et al. 2014) and active techniques which are dynamic deforming structures, such as blowing and suction (Yamamoto et al. 2013; Hasegawa and Kasagi 2011; Segawa et al. 2007; Schatzman et al. 2014; Kametani et al. 2015), plasma actuation (Duong et al. 2019, 2021; Samimy et al. 2007), and wall oscillation. Drag reduction via spanwise and wall-normal surface oscillations has been extensively studied in the literature due to its significant drag-reducing effect. In Fig. 1, we illustrate different oscillation mechanisms and compile the list of studies that have investigated those mechanisms. The maximum DR achieved so far for spanwise surface oscillation is 49% by Jung et al. (1992) (Fig. 1a), for spanwise oscillation with streamwise travelling wave is 48% by Quadrio et al. (2009). In the present study, we focus on drag reduction via wall-normal waves. For the wall-normal travelling wave mechanisms, a wall deformation propagates either in the streamwise direction (Fig. 1c) or spanwise direction (Fig. 1d). An idealized wall-normal travelling wave (TW), typically formulated as a periodic wall deformation, is given by

$$\eta(x_i, t) = \eta_{\max} \cos(\kappa_{x_i} x_i - \omega t) \quad i = 1, 3, \tag{1}$$

where $i = 1$ and $i = 3$ denote streamwise and spanwise travelling waves, respectively. $\eta_{\max}$ is the wave deformation amplitude, $\kappa_{x_i}$ is the wavenumber, and $\omega$ is the wave angular frequency. We normalize the actuation parameters with viscous scales, namely the kinematic viscosity $\nu$ and friction velocity $u_\tau$.

*Corresponding author
amir.amjadimanesh@ntu.ac.uk ( Amir Amjadimanesh)
ORCID(s):

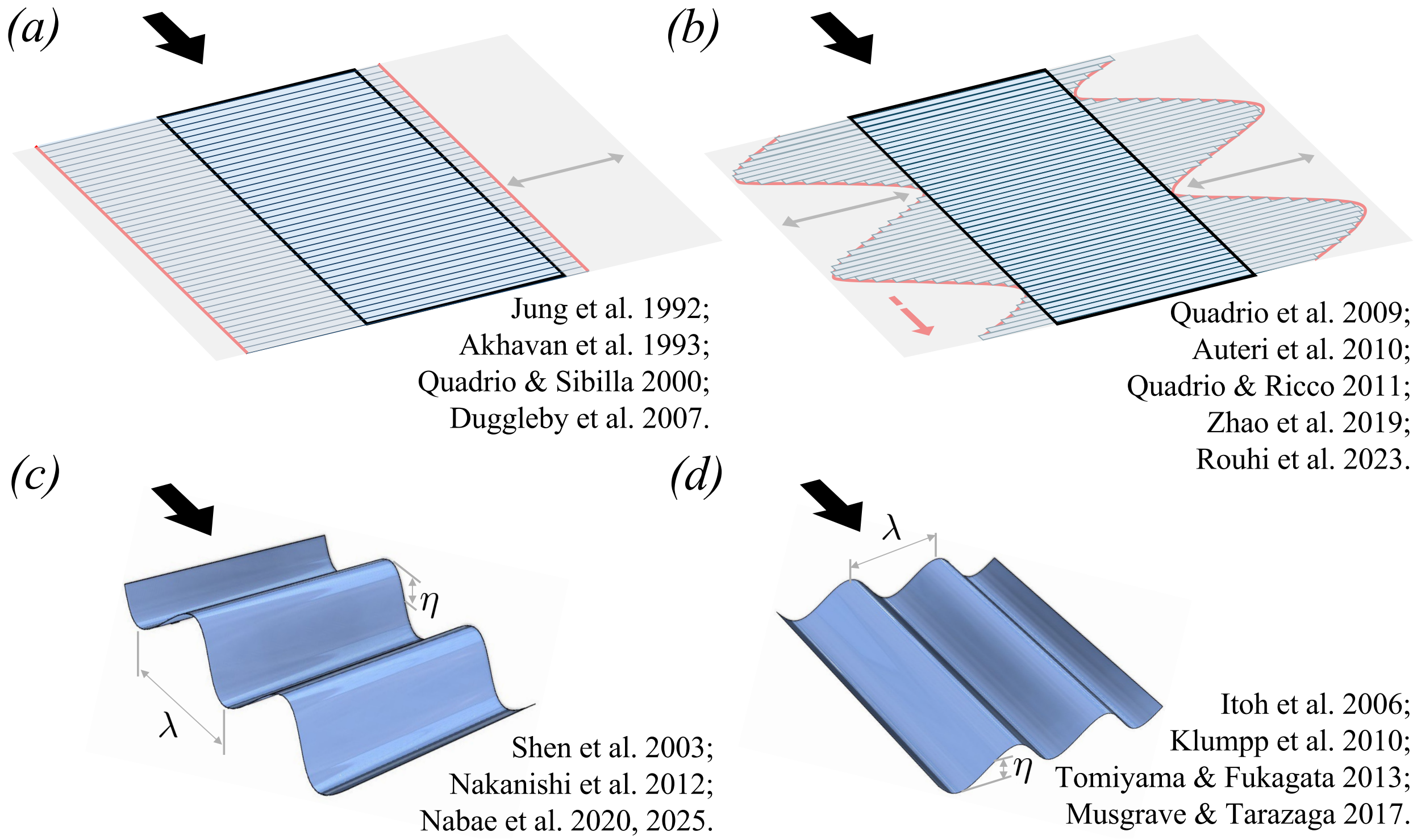

**Figure 1:** Surface oscillation and wall deformation. Spanwise (a) surface oscillation, (b) oscillation with streamwise travelling wave. Wall-normal deformation with (c) streamwise travelling wave, (d) spanwise travelling wave. The black arrow shows the flow direction.

$$\eta^+_{\max} = \frac{\eta_{\max} u_\tau}{\nu}, \quad \kappa^+ = \frac{\kappa \nu}{u_\tau}, \quad \omega^+ = \frac{\omega \nu}{u_\tau^2}$$

Then, the drag reduction ($DR$) over a wall-normal wave is defined as:

$$DR = \frac{C_{f_\circ} - C_f}{C_{f_\circ}} \times 100\ [\%], \tag{2}$$

where '$_\circ$' subscript denotes non-actuated case and $C_{f_\circ}$ and $C_f$ are the skin-friction coefficients for the non-actuated and the actuated wall, respectively, and are defined as:

$$C_f \equiv \frac{2\overline{\tau_w}}{\rho U_{b,\infty}^2},$$

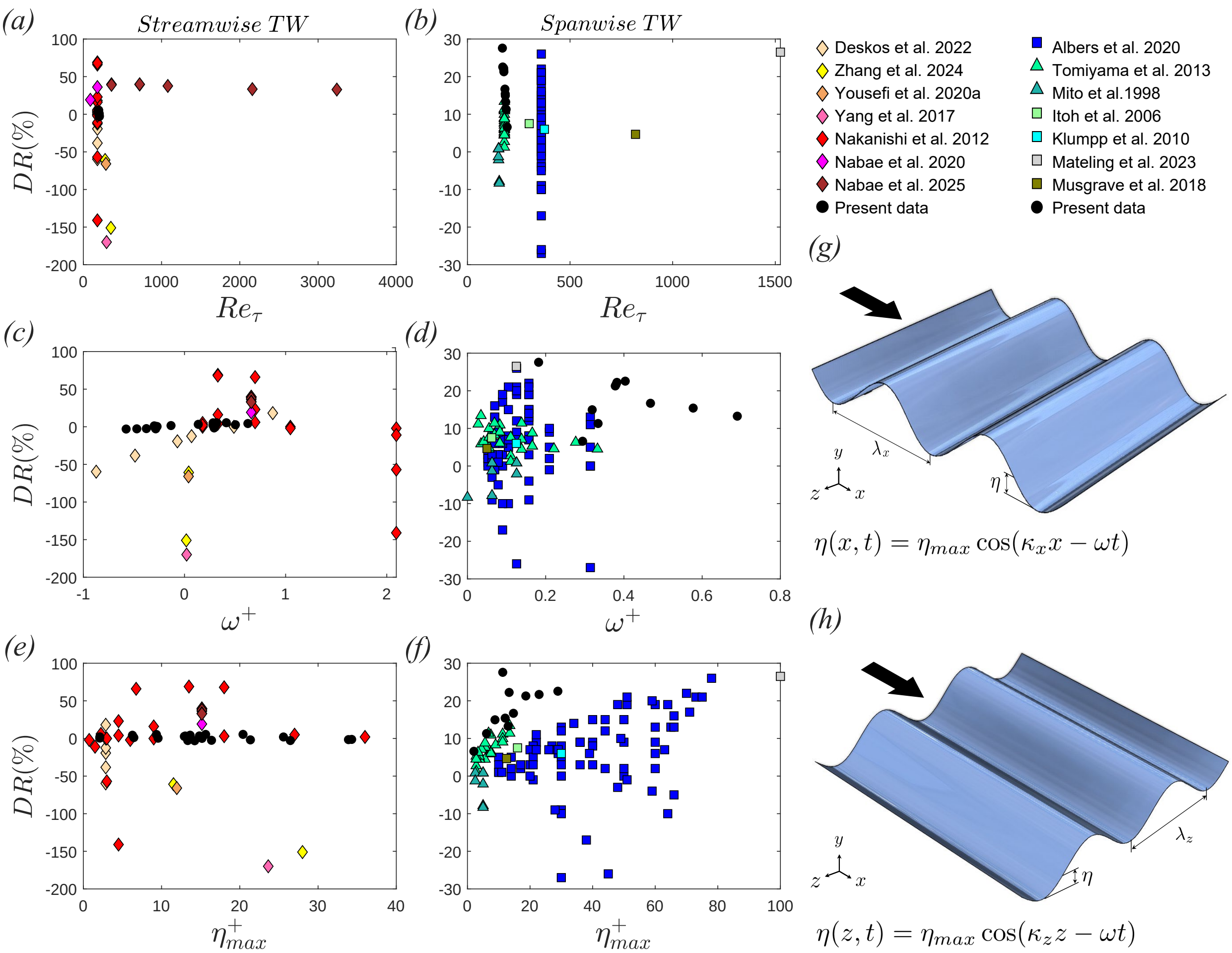

**Figure 2:** Drag reduction literature data for the wall-normal TW mechanism. Panels (a, c, e) demonstrate data points on the streamwise TWs. Panels (b, d, f) represent the data points over a spanwise TWs. Panels (a, b) display the variation of DR versus $Re_\tau$. Panels (c, d) depict the dependence of the DR on the $\omega^+$. Panels (e, f) highlight the DR against the $\eta^+_{\max}$. Panels (g, h) exhibit schematic streamwise and spanwise travelling waves, respectively. The square symbols denote boundary-layer studies, while the other symbols indicate wall-bounded flows.

where $U_{b,\infty}$ is considered as the bulk velocity for turbulent channel and pipe flows, and as free-stream velocity for turbulent boundary layer (TBL). The $\overline{\tau_w} = \rho u_\tau^2$ is the mean wall shear stress and $\rho$ is the fluid density. Hereafter, following Nabae et al. (2025), an overbar $\overline{(...)}$ denotes averaging over time and homogeneous spatial directions.

Fig. 2 displays DR against $Re_\tau$, $\omega^+$, and $\eta^+_{\max}$ for wall-normal deformation with streamwise and spanwise idealized TW. The literature includes studies of both turbulent boundary layer and turbulent wall-bounded flow using experimental setups (Yousefi et al. 2020; Itoh et al. 2006; Klumpp et al. 2010; Musgrave et al. 2018), Direct Numerical Simulation (Deskos et al. 2022; Zhang et al. 2024; Yang and Shen 2017; Nakanishi et al. 2012; Nabae et al. 2020; Tomiyama and Fukagata 2013; Mito and Kasagi 1998; Mateling et al. 2023), and Large Eddy Simulation (Nabae et al. 2025; Albers et al. 2020). Experimentalists have faced some challenges in reporting DR comprehensively because of limitations of measurement devices to adequately resolve near-wall flow characteristics (Musgrave and Tarazaga 2017). On the other hand, the study of DR at high $Re_\tau$ is computationally expensive via DNS; thus, most DNS studies have focused on $100 < Re_\tau < 500$ (Fig. 2a,b). Recent computational advances, however, have enabled LES studies of streamwise TWs at $Re_\tau$ up to 3260 (Nabae et al. 2025).

Fig. 2(a,b) show that for a fixed set of actuation parameters, DR approaches an asymptotic value at high $Re_\tau$ for both streamwise and spanwise TWs.(Albers et al. 2020; Mateling et al. 2023; Nabae et al. 2025).

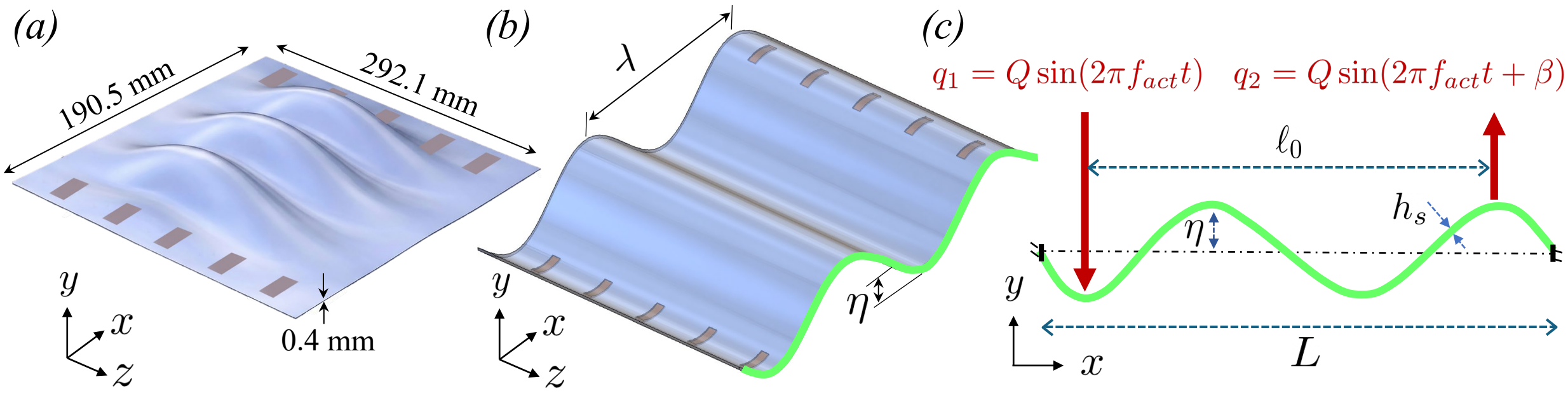

**Figure 3:** Piezoelectric actuation configuration. 3D view of AL sheet with $2 \times 5$ arrays of piezoelectric actuators beneath it for (a) Musgrave et al. (2018) setup, (b) our setup. (c) Z-view of our actuated sheet. $\eta$ is the sheet deformation. $L, h_s, D, f_{\text{nat}}$ are sheet length, thickness, flexural rigidity, and natural frequency, respectively. $l_0, \beta$ are the spacing and phase difference between actuator rows. $q_{1,2}$ are the piezoelectric actuation forces with the amplitude of $Q$. $f_{\text{act}}$ is the actuation frequency.

The variation of DR versus $\omega^+$ (Fig. 2c,d) is more dispersed. Despite many of the upstream TWs ($\omega^+ < 0$) yielding either negligible or negative DR (Zhang et al. 2024; Yousefi et al. 2020; Yang and Shen 2017), downstream TWs ($\omega^+ > 0$) lead to higher DR values in $0.2 \leq \omega^+ \leq 1$. While studies on spanwise TWs focus on $0.1 < \omega^+ < 1$, within which DR peaks at 26% around $\omega^+ \approx 0.25$ (Albers et al. 2020; Mateling et al. 2023).

In terms of $\eta^+_{\text{max}}$ (Fig. 2e,f), the maximum DR for streamwise TWs corresponds to $10 \leq \eta^+_{\text{max}} \leq 20$ (Nakanishi et al. 2012; Nabae et al. 2020, 2025). Nevertheless, the DR datapoints for spanwise TWs show considerable dispersion with $\eta^+_{\text{max}}$. The maximum achievable drag reduction converges to 30%, revealing that DR is less sensitive to spanwise TW parameters.

Previous numerical studies have applied an idealised travelling wave (1) as the bottom-wall boundary condition for turbulent channel or TBL. However, in practice, we require a mechanism that mimics (1). Musgrave et al. (2018) used piezoelectric actuation as a potential mechanism (Fig. 3a). They placed piezoelectric actuators beneath a thin Aluminium sheet. The actuators were driven by an alternative voltage of $\pm 350\ V$, producing deformations with $0.04 \leq \omega^+ \leq 0.1$ and $5.9 \leq \eta^+_{\text{max}} \leq 12.5$ (corresponding to $350Hz \leq f \leq 810Hz$ and $117\mu m \leq \eta^+_{\text{max}} \leq 249\mu m$). To conduct wind-tunnel experiments at $Re_\tau \approx 1000$, the sheet was fully clamped, and measurements were performed using hot-wire anemometry. Although the actuation system successfully produced surface deformations, full clamping and wave reflections at the sheet boundaries resulted in unintended three-dimensional, complex deformation patterns. Therefore, from a practical point of view, realistic waves deviate from idealised waves, hence Musgrave et al. (2018) could not produce 2D waves. While Musgrave et al. (2018) achieved drag reduction, the limited near-wall resolution of hot-wire anemometry prevented a quantitative and comprehensive assessment of the drag reduction.

Here, we restrict ourselves to unilateral surface deformation, enabling us to systematically study the streamwise and spanwise wall-normal waves. To our knowledge, this is the first computational study to consider surface deformations using a setup that could be implemented in practice, namely, piezoelectric actuation. Through DNS, we are able to 1) investigate different sets of actuation parameters, 2) accurately calculate DR, and 3) investigate the flow physics to an unprecedented detail. We pursue these objectives in this study. We simulate our piezoelectric structural system using Finite Element Analysis (FEA). Then, we apply the resulting waveform function as the bottom-wall boundary condition of a turbulent half-channel flow with the half-channel height of $h$. We solve the governing flow equations using a well-validated DNS setup. We conduct computations at a constant flow rate with $Re_b = 2hU_{b,\infty}/\nu = 6312$ ($Re_\tau = hu_\tau/\nu \approx 200$), and study the actuation parameter space ($2 \leq \eta^+_{\text{max}} \leq 34$, and $-0.58 \leq \omega^+ \leq 0.70$) to identify configurations that maximise DR (§2). We achieved a maximum drag reduction of 5.4% with streamwise piezo-generated waves at $\omega^+ = 0.3$ and $\eta^+_{\text{max}} = 20.4$. The streamwise waves induce a local adverse/favourable pressure gradient in their propagation direction, resulting in a change in drag. The maximum achievable DR for the spanwise waves is 27.6% at $\omega^+ = 0.17$ and $\eta^+_{\text{max}} = 11.5$. The wave induces the spanwise wind, leading to attenuated Reynolds stresses and the clustering of high- and low-speed streaks (§3). Finally, the study's key findings are summarised in §4.

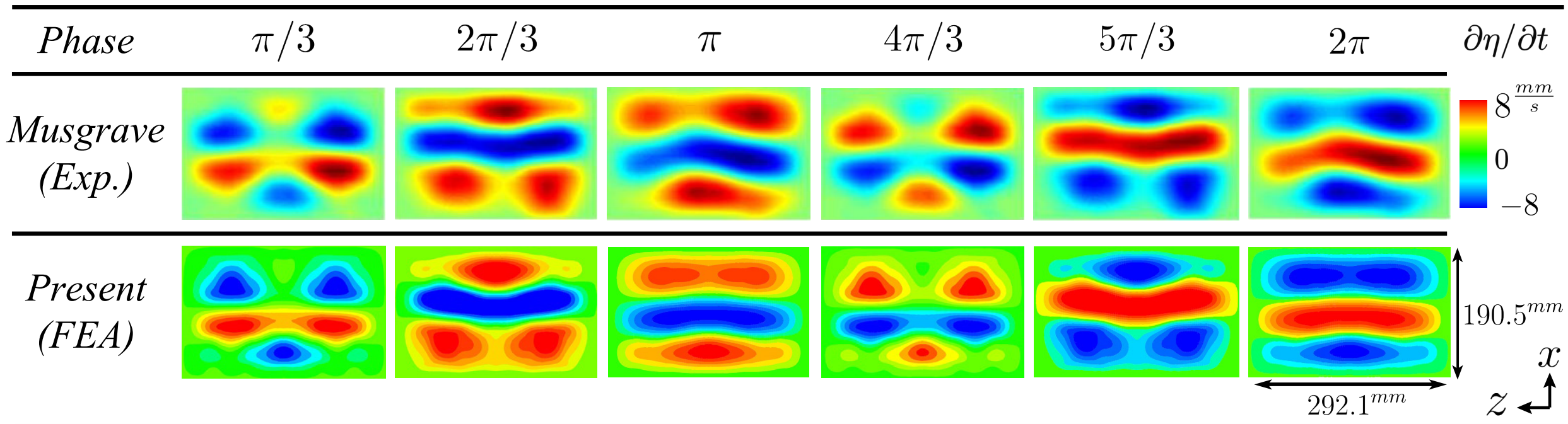

**Figure 4:** Al sheet's rate of deformation contours of the experimental study by Musgrave et al. (2018) and the present FEA at six actuation phases. There is a phase difference of $100^o$ between the top and bottom actuator rows. The actuation voltage and frequency are $\pm350V$ and $310Hz$, respectively.

## 2. Methodology

### 2.1. Piezoelectric setup

We employed FEA in ANSYS® Mechanical to simulate piezoelectric actuation and the resulting sheet deformation. To validate our FEA setup, we replicated the configuration of Musgrave and Tarazaga (2017) (Fig. 3a), consisting of a fully clamped aluminum sheet actuated by two rows ($2 \times 5$) of MFC4312-P1 piezoelectric patches. A phase difference of 100° was imposed between the actuator rows. The actuators were driven by an alternative voltage (AC) of $\pm350$ $V$ at 310Hz. Fig. 4 compares our deformation rate at six phases with Musgrave et al.'s experimental data. Our FEA results agree well with experimental ones, verifying the accuracy of our setup.

The setup by Musgrave et al. generates complex deformations $\eta(x, z, t)$ owing to the fully clamped, low-aspect-ratio aluminum sheet. However, for our production DNSs, we only clamp the two opposite sides that are adjacent to the piezoelectric rows (Fig. 3b,c), hence generating deformations along only one direction, e.g. $\eta(x, t)$.

Fig. 3(c) illustrates our setup with the actuation parameters involved. Each actuator is connected to an AC voltage, and exerts a sinusoidal force on the sheet with length $L$. We introduce a phase difference $\beta$ between the actuator rows. We set the phase difference ($\beta$) based on the row spacing ($l_0$) and the wavelength ($\lambda$), to eliminate wave reflection at the clamped boundaries (Anakok et al. 2022; Tomikawa et al. 1990).

$$\beta = \frac{2\pi l_0}{\lambda}$$

The Aluminum sheet under our considered actuation parameters falls into the linear elastic deformation regime. In the absence of mechanical losses, the deformation is governed by the Thin Plate Theory (Timoshenko and Woinowsky-Krieger 1959; Landau and Lifshitz 1986):

$$D\,\frac{\partial^4\eta(x,t)}{\partial x^4} + \rho_s h_s\,\frac{\partial^2\eta(x,t)}{\partial t^2} = q. \tag{3}$$

Here $\rho_s$, $h_s$, $D$ are the sheet density, thickness, and flexural rigidity, respectively. $q$ is the force per unit area exerted by the piezoelectrics.

Considering the parameters involved in (3) and the geometrical parameters (Fig. 3c), the maximum sheet deformation is a function of seven variables,

$$\eta_{\max} = f'(L, h_s, \rho_s, D, l_0, Q, f_{\text{nat}}, f_{\text{act}}), \tag{4}$$

where $f_{\text{act}}$ denotes the actuation frequency of piezoelectrics, and $Q$ represents the piezo force amplitude per unit area. To arrive at a unified non-dimensional parameter space and reduce the number of independent variables, we apply Buckingham $\Pi$ theorem, which, after some recasting, yields:

$$\frac{\eta_{\max}}{L} = f\left(\frac{l_0}{L}, \frac{h_s}{L}, G_i^2\frac{f_{\text{act}}}{f_{\text{nat}}}, \frac{QL^3}{D}\right), \tag{5}$$

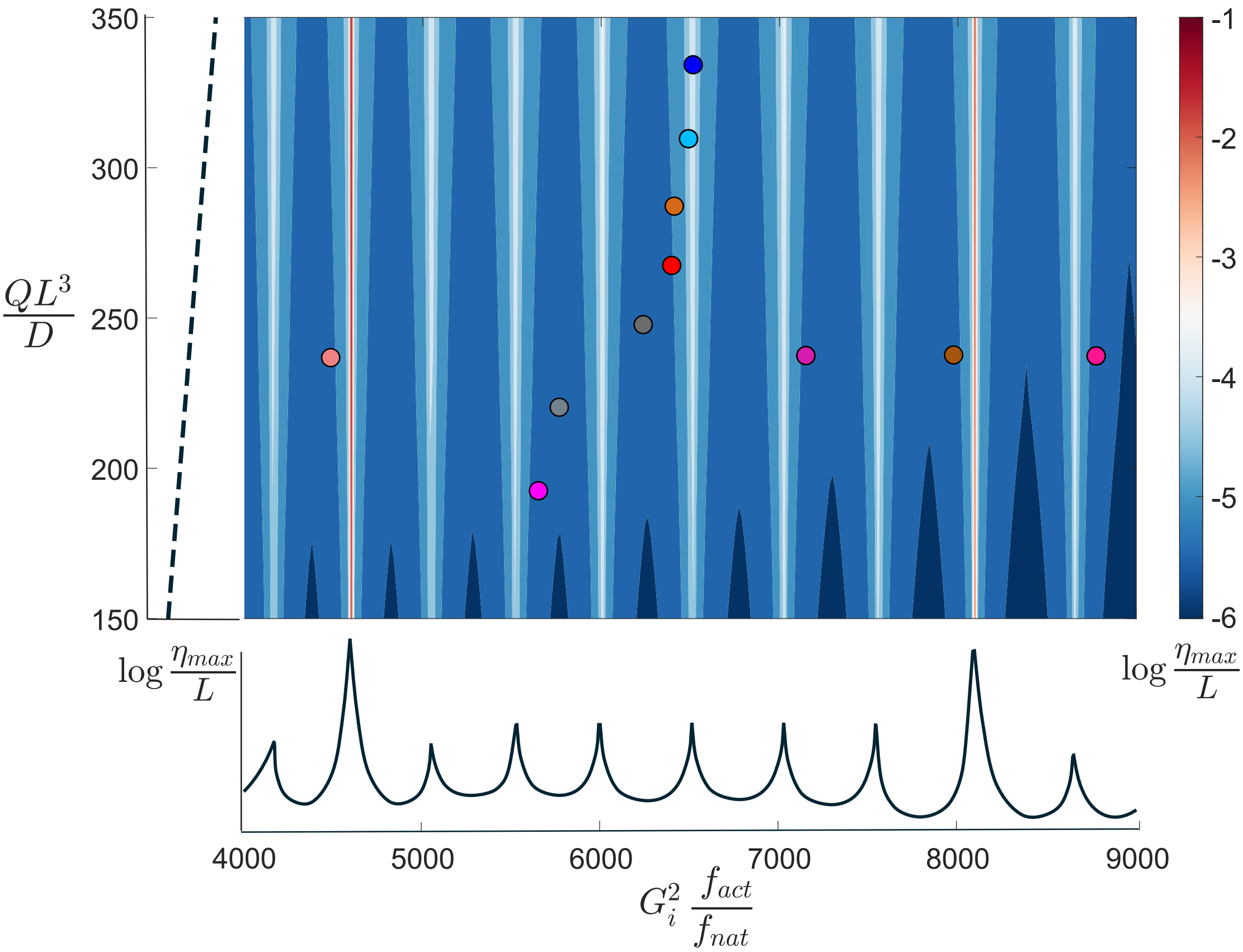

**Figure 5:** Piezoelectric actuation map. Variation of the wall deformation against actuation frequency and voltage force as the abscissa and ordinate, respectively (all in dimensionless form). Gray, blue, and red spectra bullets denote the TW, SW, and HW, respectively. The black solid and dashed graphs denote the variation of $log(\eta_{\max}/L)$ with $G_i^2 f_{\text{act}}/f_{\text{nat}}$ and $QL^3/D$, respectively.

where $f_{\text{nat}}$ denotes the system's natural frequency and $G_i^2$ is the dimensionless frequency parameter of the $i^{th}$ vibration mode of the sheet, dependent on the boundary condition (Leissa 1993). For our setup, the material properties of the Aluminum sheet, as well as its dimensions ($l_0 = 600mm$, $L = 686mm$, $h_s = 0.4mm$), are fixed, so $l_0/L$ and $h_s/L$ are fixed as well. Therefore, the varying parameters in (5) are the last two arguments associated with the actuation frequency $f_{\text{act}}$ and the forcing amplitude $Q$. We sweep over practical ranges of piezoelectric frequency and voltage (force amplitude) for our system. We plot the $\log(\eta_{max}/L)$ against $G_i^2 f_{\text{act}}/f_{\text{nat}}$ and $QL^3/D$ in Fig. 5. The deformation map varies intermittently with $f_{\text{act}}$, but grows exponentially with Q, $\log(\eta_{max}/L) \propto Q$ hence $\eta_{max} \propto \exp(Q)$.

The actuation parameters not only affect the deformation amplitude but also control the wave type, ranging from a pure travelling wave to a pure standing wave. Fig. 6 illustrates the emerging wave types, which could be categorized into pure travelling (TW, 1), pure standing waves (SW), and hybrid (HW).

The resulting deformation (Fig. 5) from actuation setup (Fig. 3c) is dominated by the first two modes. Thus, $\eta$ could be expressed as the superposition of two cosine waves with identical wavenumbers and frequencies, but with a phase lag,

$$\eta(x,t) = \eta_1 \cos(\kappa x - \omega t + \phi_1) + \eta_2 \cos(\kappa x - (-1)^n \omega t + \phi_2), \tag{6}$$

when $n = 0$ and $2\phi_1 = -\phi_2$, (6) yields TW (Fig. 6, left), when $n = 1$ and $\phi_1 = -\phi_2$, (6) yields SW (Fig. 6, right), and when $n = 1$ and $\phi_1 - \phi_2 = \pi/2$, (6) yields HW (Fig. 6, middle).

Negative values of $\omega$ correspond to upstream travelling waves. The superposition in (6) can be reduced to a single cosine function with an effective amplitude

$$\eta_{\text{eff}} = [\eta_1^2 + 2\eta_1\eta_2 \cos(\phi_1 - \phi_2 - 2n\omega t) + \eta_2^2]^{1/2}.$$

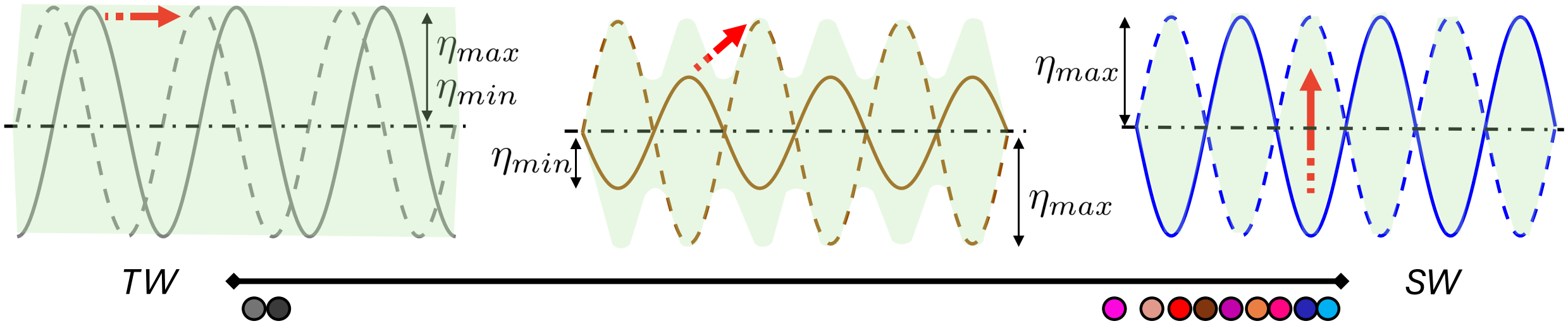

**Figure 6:** Schematic of wave types and envelopes. From Left to right: travelling waves to hybrid and standing waves. Green-shaded area presents the wave envelope.

The maximum attainable amplitude can be simplified to:

$$\eta_{\max} = [\eta_1^2 + 2\eta_1\eta_2\cos(3\phi_1) + \eta_2^2]^{1/2}, \qquad for\ TW, \tag{7a}$$

$$\eta_{\max} = \eta_1 + \eta_2, \qquad for\ HW, \tag{7b}$$

$$\eta_{\max} = \eta_1 + \eta_2, \qquad for\ SW. \tag{7c}$$

Fig. 5 reveals that actuation frequencies lying exactly between two adjacent harmonics of $f_{\text{nat}}$ produce a travelling waveform (1), characterized by a minimum deformation amplitude and a linear wave envelope (gray bullets, Fig. 6 Left drawing). In contrast, when the $f_{\text{act}}$ matches one of the harmonics of $f_{\text{nat}}$, the result is a standing waveform with maximum deformation amplitude and a sinusoidal wave envelope (blue bullets, Fig. 6 Mid drawing). When $f_{\text{act}}$ falls between these two extremes, it generates a combination of travelling and standing waves, referred to as a hybrid wave, with an intermediate deformation amplitude (red spectrum bullets, Fig. 6 right drawing). It is also worth noting that the travelling direction of the hybrid wave reverses after crossing the adjacent travelling wave frequency. Overall, the actuation map highlights the potential of piezoelectric actuation to generate diverse surface deformations.

## 2.2. Numerical flow set-up

We solve conservation of mass and momentum for an incompressible fluid in a turbulent half-channel flow:

$$\nabla \cdot \mathbf{u} = 0, \tag{8a}$$

$$\frac{\partial \mathbf{u}}{\partial t} + \boldsymbol{\nabla} \cdot (\mathbf{uu}) = -\frac{1}{\rho}\nabla p + \nu\nabla^2\mathbf{u} - \frac{1}{\rho}\frac{dP}{dx}\hat{e}_x, \tag{8b}$$

where $\mathbf{u} = (u, v, w)$ is the velocity vector in the streamwise, wall-normal, and spanwise directions $(x, y, z)$, respectively. The total pressure gradient is decomposed into the periodic part $(\nabla p)$ and the driving part $(dP/dx)$. Equations (8a, 8b) are transformed into the curvilinear coordinate system $(\xi, \zeta, z)$, and solved using an in-house boundary-fitted, pseudo-spectral DNS solver. Spatial discretization is the Fourier-Galerkin method in the streamwise and spanwise directions, while a central finite-difference scheme is applied in the wall-normal direction. A hybrid time-integration scheme is employed; all the terms are integrated using a three-step second-order Runge-Kutta scheme, except for the wall-normal diffusion term, which is integrated using a second-order Crank-Nicolson scheme. Fractional-step algorithm (Chorin 1969, Kim and Moin 1985) is used to march (8a) and (8b). We apply periodic boundary conditions in the streamwise and spanwise directions, and no-slip (with travelling-wave kinematics) and free-slip conditions at the bottom and top boundaries, respectively.

We have extensively verified the accuracy of our DNS solver against benchmark wall-bounded turbulent flows related to our production runs. In Appendix A, we present our validation results for smooth-wall turbulent channel flow at $Re_\tau = 395$, as well as turbulent channel flow over a wavy wall at $Re_\tau = 215$. In addition, the present travelling-wave implementation is validated against the slow and fast travelling wave cases by Zhang et al. (2024)'s study. The computational configuration is shown in Fig. 7(a), with simulation parameters summarised in Table 1. To ensure consistency, we matched both the open-channel dimensions and $Re_\tau$ of each reference case. We applied a streamwise idealistic travelling wave (1) as the bottom-wall boundary condition. While Zhang et al. (2024) solved the governing equations using OpenFOAM (Weller et al. 1998), we employed our in-house solver MELA.

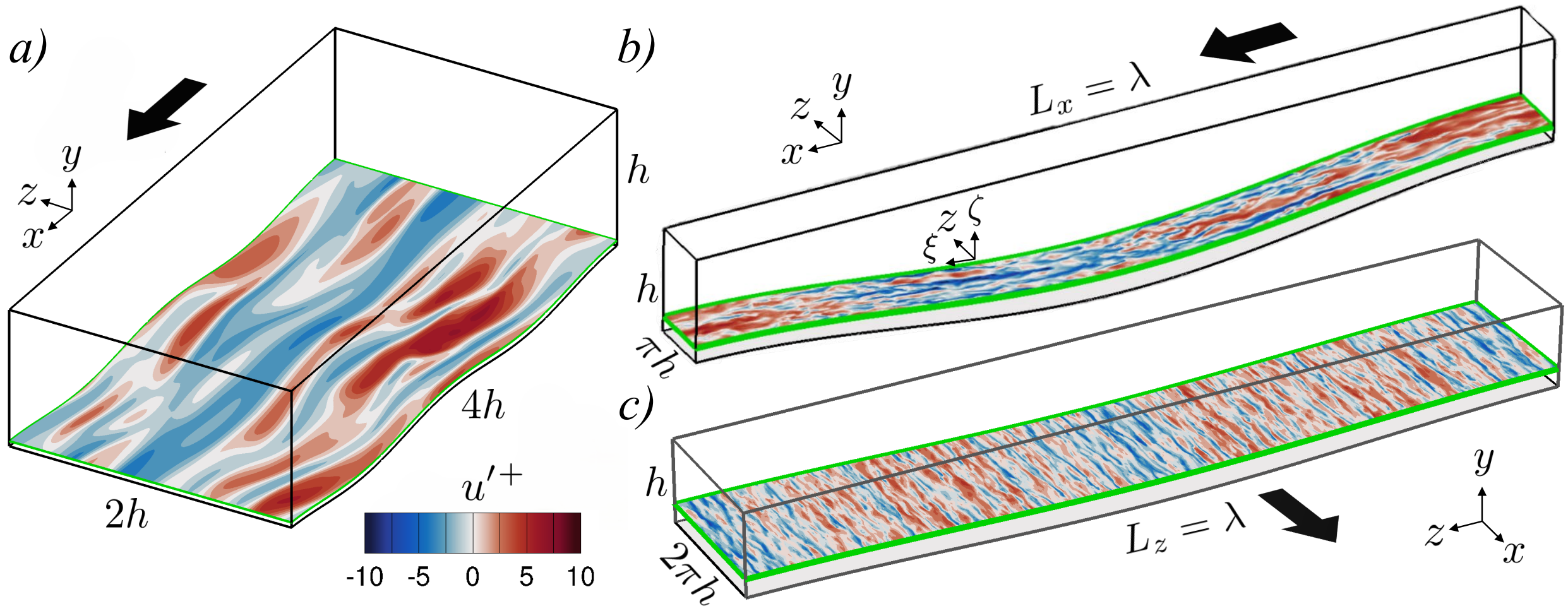

**Figure 7:** Computational domains for (a) Zhang et al. (2024)'s fast case, (b) streamwise wave case of the present study, and (c) spanwise wave case of the present study. The black arrow indicates the flow direction. The contour field plots the streamwise fluctuating velocity ($u'^+$) at 15 viscous units away from the wall.

| Case | $\eta^+_{\max}$ | $\omega^+$ | $\kappa^+$ | $c^+$ | Domain size | $Re_\tau$ | $Re_b$ | Grid | $\Delta\xi^+$ | $\Delta\zeta^+_w$ | $\Delta z^+$ |
|---|---|---|---|---|---|---|---|---|---|---|---|
| Slow | 12.03 | 0.04 | $1.08e{-}2$ | 3.69 | $2\lambda_x \times 0.5\lambda_x \times \lambda_x$ | 278.5 | 7300 | Coarse | 17.4 | 0.9 | 8.7 |
| | | | | | | | | Fine | 8.7 | 0.4 | 4.3 |
| | | | | | | | | Ref. | 4.5 | 0.6 | 4.5 |
| Fast | 8.96 | 0.25 | $1.45e{-}2$ | 17.25 | $2\lambda_x \times 0.5\lambda_x \times \lambda_x$ | 211 | 7300 | Coarse | 13.2 | 0.7 | 6.6 |
| | | | | | | | | Fine | 8.8 | 0.4 | 4.4 |
| | | | | | | | | Ref. | 3.4 | 0.4 | 3.4 |

**Table 1**
Details of travelling wave studies of Zhang et al. (2024) (Ref.) with surface deformation are in equation (1). The cases denote slow and fast wave speeds ($c^+ = \omega^+/\kappa^+$), respectively. The present coarse and fine grid resolutions are summarized for comparison with the reference.

Fig. 8 compares the profiles of the streamwise mean velocity ($\overline{u}^+ = \overline{u}/u_\tau$) and Reynolds shear stress ($\overline{u''v''}^+$) between our cases and those by Zhang et al. (2024). The $\overline{u}^+$ profiles from our coarse and fine grids are in good agreement with those by Zhang et al. (2024) (Fig. 8a,c). Slight differences are seen for the slow travelling wave case near the wake region ($\zeta^+ \gtrsim 100$). This relates to the smaller domain size reported by Zhang et al. (2024). Truncation of the domain width below $L_z \simeq \pi$, under-resolves the outer region of turbulent channel flow (Lozano-Durán and Jiménez 2014; Jiménez 2018).

We calculate $\overline{u''v''}^+$ in Fig. 8(b,d), using triple decomposition (Hussain and Reynolds 1970):

$$f(\xi,\zeta,z,t) = \underbrace{\overline{f}(\zeta) + \tilde{f}(\xi,\zeta,t)}_{\langle f(\xi,\zeta,t)\rangle} + f''(\xi,\zeta,z,t), \tag{9a}$$

$$\langle f(\xi,\zeta,t)\rangle = \frac{1}{M}\sum_{m=0}^{M-1}\frac{1}{L_z}\int_0^{L_z} f(\xi,\zeta,z,t+mT_{\mathrm{osc}})\,\mathrm{d}z, \tag{9b}$$

where $\overline{f}(\zeta)$ is the mean part, averaged over $\xi$, $z$ and $t$, $\langle f(\xi,\zeta,t)\rangle$ is the spanwise and phase-averaged part, $\tilde{f} = \langle f\rangle - \overline{f}$ is the harmonic part, and $f'' = f - \langle f\rangle$ is the fluctuating (turbulent) part.

In terms of $\overline{u''v''}^+$ profiles (Fig. 8b,d), both our coarse and fine grids agree well with the profile of Zhang et al. (2024) for the slow TW case (Fig. 8b). For the fast TW case, we need a fine grid resolution to achieve agreement with the profile in Zhang et al. (2024) (Fig. 8d).

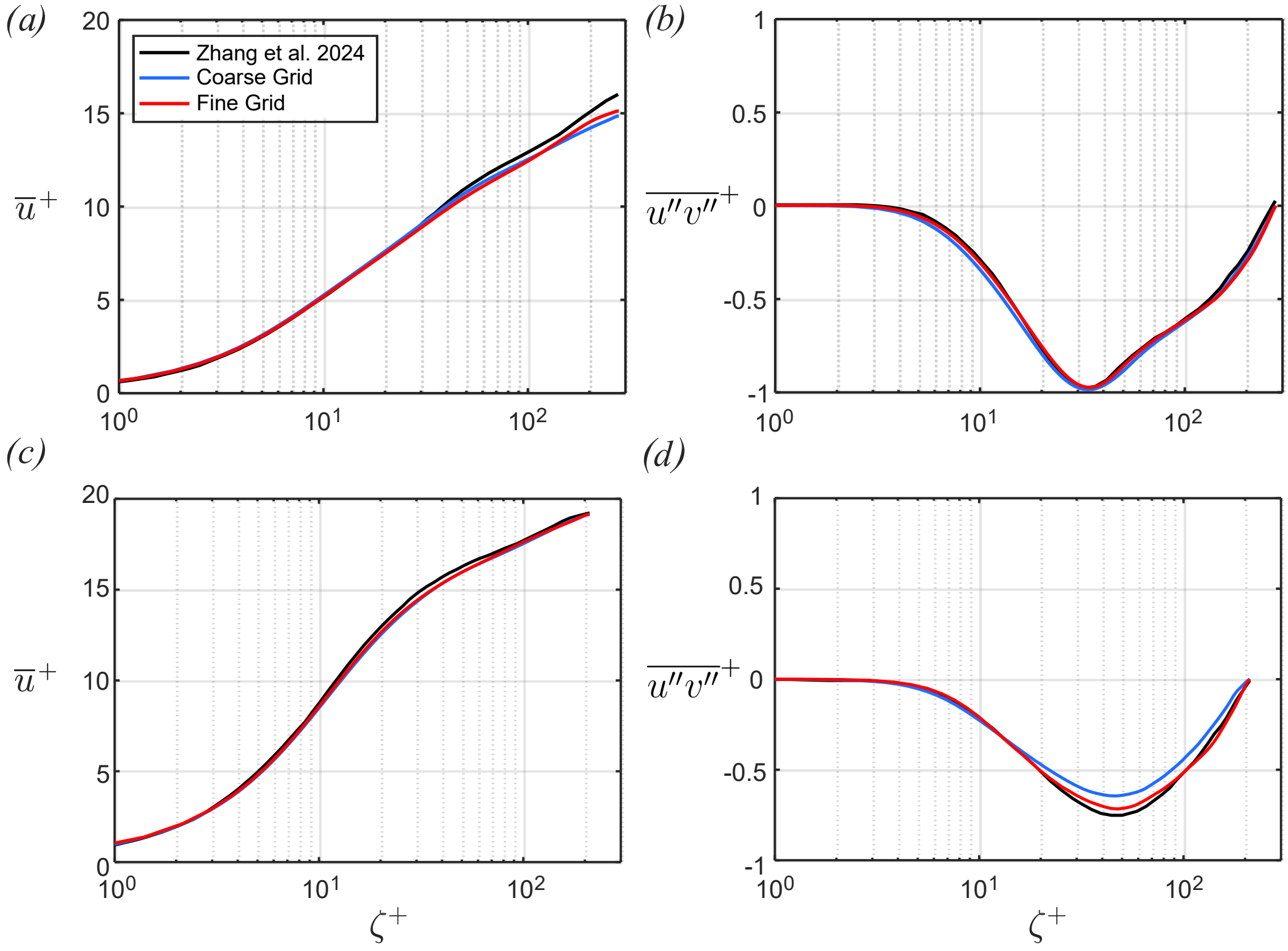

**Figure 8:** Comparison of the streamwise mean velocity profiles (a,c), and Reynolds shear-stress profiles (b,d) from the validation cases of Table 1. Panels (a,b) present the slow travelling wave case, while panels (c,d) demonstrate the fast travelling wave case.

## 2.3. Simulation cases

Dimensional analysis for the drag reduction of wall-normal deformation yields:

$$DR = g(\kappa^+, \omega^+, \eta^+_{\max}, Re_\tau), \tag{10}$$

reviewing (7), $\eta^+_{\max} = \eta^+_{\max}(\eta_1, \eta_2, \phi_1)$. The first three dimensionless groups in (10) are the wave characteristics, associated with the actutation map (Fig. 5).

Fig. 7(b,c) illustrate the computational domain for the streamwise and spanwise wave setup, respectively. For our production simulations, we sweep over actuation parameters in the practical ranges of piezo voltages $250V \leq Q \leq 500V$ and frequencies $119Hz \leq f_{\text{act}} \leq 543Hz$. To convert our actuation parameters to non-dimensional viscous units, we assume that the piezoelectric setup is implemented in the experimental pipe-flow setup by Ding et al. (2024). For conversion, we consider their operating condition, namely their $u_\tau$ and $\nu$, at $Re_\tau = 1356$. However, to afford for a parametric study via DNS, we conduct our calculations at $Re_\tau = 200$. The flow physics and turbulence statistics over active and passive controlling mechanisms could be extrapolated from low to high Reynolds numbers, provided that the actuation parameters are fixed in viscous units (Marusic et al. 2013; Quadrio and Ricco 2004). The surface deformation and the actuation frequency in viscous units vary within the range $2 \leq \eta^+_{\max} \leq 34$, and $-0.58 \leq \omega^+ \leq 0.70$.

We select 11 cases for each downstream, upstream, and spanwise waves, covering the three wave types (TW, SW, and HW). These 11 cases are marked on the actuation map (Fig. 5). In Appendix B (Table 5), we provide details of our production calculations, including the actuation parameters, domain and grid sizes, and drag reduction.

We apply the wave functions as the bottom-wall boundary condition of the half-channel. For all cases, we drive the flow in the x-direction. For the DNSs over the streamwise wave deformation, the wave propagates in the x-direction (Fig. 7b), and for the DNSs over the spanwise wave deformation, the wave propagates in the z-direction (Fig. 7c).

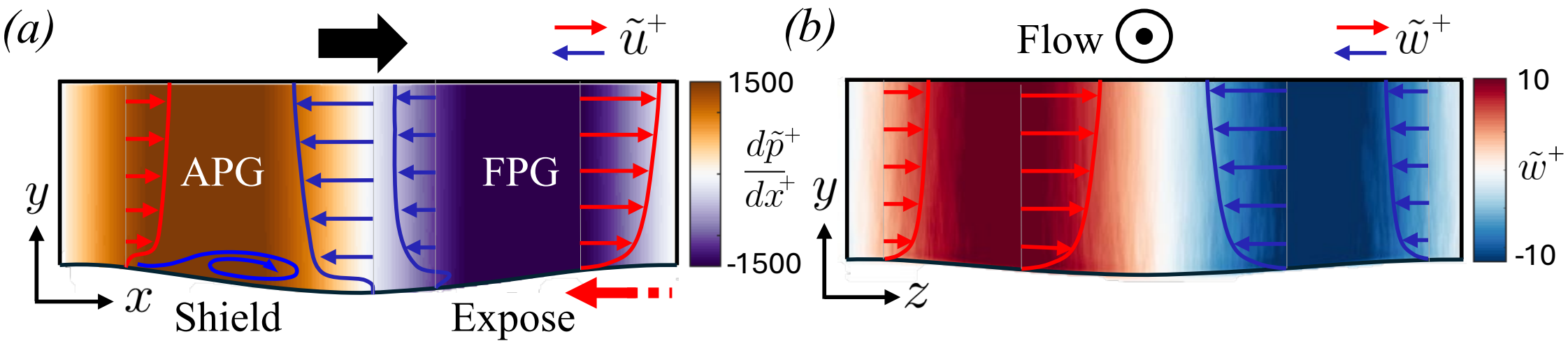

**Figure 9:** Schematic illustration of the physical mechanisms underlying drag reduction for (a) upstream, and (b) spanwise waves. Panel (a) shows the side view of the harmonic streamwise velocity profiles ($\tilde{u}^+$) and harmonic pressure gradient ($d\tilde{p}^+/dx^+$). Panel (b) presents a front view of the spanwise-wave case, including the harmonic spanwise velocity profiles ($\tilde{w}^+$).

## 3. Results and discussions

### 3.1. Overall flow physics

To aid discussion throughout the results section, we schematically illustrate the overall flow physics in Fig. 9. Surface deformation locally expands and contracts the channel cross-sectional area. As a result, the phase-averaged flow consists of an oscillating pressure gradient and an oscillating wind, both of which interact with the near-wall turbulence. In the case of a streamwise wave (Fig. 9a), the near-wall turbulence is exposed to alternating favorable and adverse pressure gradients (FPG and APG). The FPG attenuates the near-wall turbulence, while the APG augments it. In the case of a spanwise wave (Fig. 9b), the oscillating pressure gradient (wind) is perpendicular to the advection direction of the near-wall turbulence. As a result, the near-wall turbulence is exposed to an oscillating transverse shear. Such interaction resembles turbulent flows over spanwise plane oscillation (Touber and Leschziner 2012), and spanwise oscillation with streamwise traveling wave (Rouhi et al. 2025). In these cases, the wall oscillation generates a Stokes layer, which, in turn, imposes an oscillating spanwise shear on the near-wall turbulence.

### 3.2. Overall variations of DR

Considering (10), DR is a function of four dimensionless groups. Several studies have reported that DR induced by wall-normal deformation is governed primarily by the wave frequency (or phase speed) and amplitude, while DR does not change significantly with wavenumbers over the range $0.015 < \kappa^+ < 0.035$ (Nabae et al. 2025; Ricco et al. 2021; Tomiyama and Fukagata 2013; Itoh et al. 2006). In the present study, the wavenumbers are sufficiently small ($0.0005 < \kappa^+ < 0.001$, $6000 < \lambda^+ < 12000$), that means the wall deformation varies over length scales far exceeding those of the near-wall streaks ($\lambda^+_{streak} \simeq 1000$, Jiménez and Moin 1991; Smith and Metzler 1983). Therefore, we focus on the effect of $\omega^+$ and $\eta^+_{\max}$ on DR (Fig. 10). Consistent with Fig. 5, the gray bullets indicate travelling waves, the blue bullets denote standing waves, and the red spectrum bullets are related to hybrid waves. In agreement with the literature (Zhang et al. 2024; Yousefi et al. 2020; Yang and Shen 2017), upstream waves generally result in either a drag increase or a slight drag reduction (Fig. 10a). The maximum DR for the upstream waves corresponds to $(\eta^+_{\max}, \omega^+) = (13, -0.13)$. The maximum DR for the downstream waves is about 6% with hybrid waves at $(\eta^+_{\max}, \omega^+) \approx (15, 0.4)$ (Fig. 10b), which is consistent with the literature (Nakanishi et al. 2012; Nabae et al. 2020, 2025). We achieve significantly higher DRs with the spanwise waves (Fig. 10c). The maximum achievable DR is about 27% for a hybrid wave at $(\eta^+_{\max}, \omega^+) \approx (12, 0.2)$.

The overall trend of the DR with wave amplitudes indicates that the optimal value of $\eta^+_{\max}$ is 10 to 15 for both streamwise and spanwise waves.

In Fig. 10(d,e,f), we plot the DR data versus $\eta^+_{\max}\omega^+$, which is the amplitude of wall-deformation velocity. We observe a clearer trend in DR versus $\eta^+_{\max}\omega^+$, compared to DR versus $\omega^+$ (Fig. 10a,b,c). This is due to the direct relation between the oscillating wind velocity ($\tilde{u}$ for the streamwise wave, and $\tilde{w}$ for the spanwise wave), as will be discussed in §3.4.

For the streamwise travelling wave (Fig. 10d,e), DR increases from -3% at $\eta^+_{\max}\omega^+ \simeq 10$ to about 6% by $\eta^+_{\max}\omega^+ \simeq 7$. Increasing $\eta^+_{\max}\omega^+$ beyond 7, leads to a drop in DR. For the spanwise travelling wave (Fig. 10f), DR increases with $\eta^+_{\max}\omega^+$ up to its highest value that we have simulated.

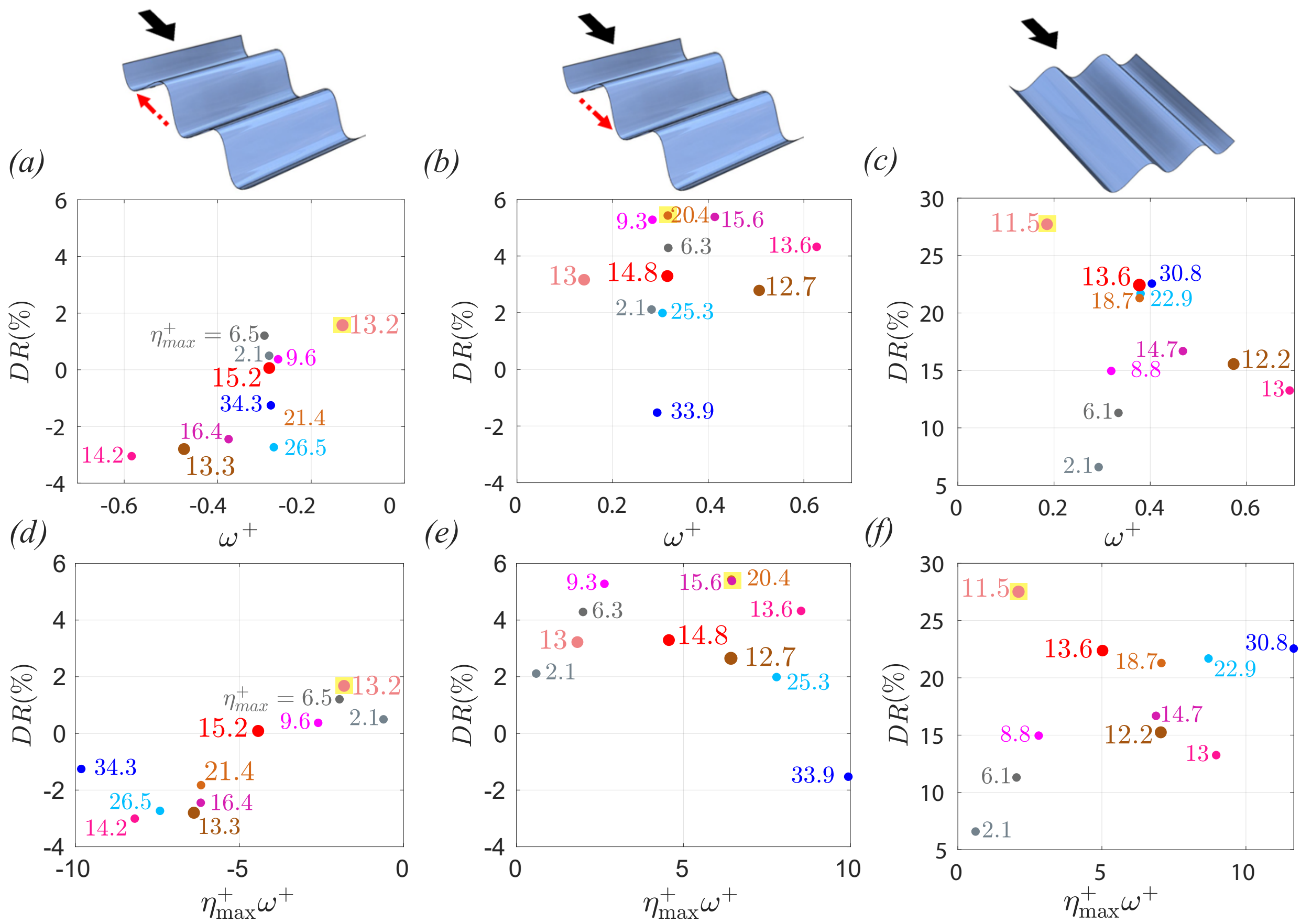

**Figure 10:** DR of case studies. DR versus $\omega^+$ for the (a,b,c) upstream waves, downstream waves, and spanwise waves, respectively. DR versus $\eta^+_{\max}\omega^+$ for the (d,e,f) upstream waves, downstream waves, and spanwise waves, respectively. Blue, gray, and red bullets denote standing, travelling, and hybrid waves, respectively. The numbers next to the bullets correspond to the wave amplitude ($\eta^+_{\max}$). The highlighted bullets are associated with the maximum DR.

| Wave ($CF[\%]$) | $\kappa^+ \times 10^4$ | $\omega^+$ | $\eta^+_{\max}$ | $\eta^+_{\max}\omega^+$ | $DR[\%]$ | Profile Color |
|---|---|---|---|---|---|---|
| USHW-90 | 4.95 | −0.13 | 13 | -1.7 | 1.6 | Pink |
| USHW-91 | 7.36 | −0.29 | 14.8 | -4.3 | 0.1 | Red |
| USHW-92 | 9.30 | −0.47 | 12.7 | -6 | -2.7 | Brown |
| DSHW-90 | 5.01 | 0.14 | 13.2 | 1.8 | 3.2 | Pink |
| DSHW-91 | 7.58 | 0.31 | 15.2 | 4.7 | 3.4 | Red |
| DSHW-92 | 9.61 | 0.50 | 13.3 | 6.7 | 2.7 | Brown |
| SPHW-90 | 5.80 | 0.18 | 11.5 | 2.1 | 27.6 | Pink |
| SPHW-93 | 8.34 | 0.38 | 13.6 | 5.2 | 21.3 | Red |
| SPHW-92 | 10.30 | 0.57 | 12.2 | 6.9 | 15.4 | Brown |

**Table 2**
Summary of upstream, downstream, and spanwise delegate wave cases and corresponding profile colors.

## 3.3. Mean velocity profiles

To explain the trends in DR, we select three cases for upstream, downstream, and spanwise waves (Table 2). These cases have actuation parameters within the ranges $0.1 < \omega^+ < 0.6$ and $11 < \eta^+_{\max} < 16$. Hereafter, these cases are distinguished by different profile colors (*Pink:* $\omega^+\eta^+_{\max} \approx 2$, *Red:* $\omega^+\eta^+_{\max} \approx 5$, *Brown:* $\omega^+\eta^+_{\max} \approx 7$).

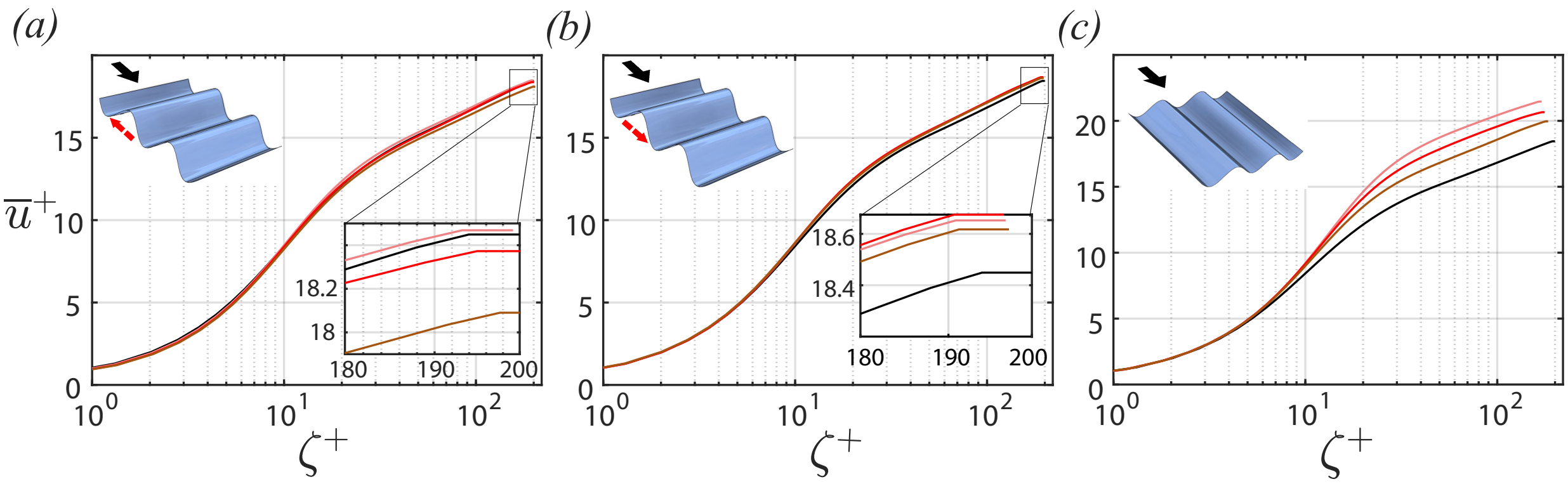

**Figure 11:** Streamwise mean velocity profiles for (a) upstream waves, (b) downstream waves, and (c) spanwise waves. Profile colors transition from pink to red to brown, representing increasing values of $\eta^+_{\max}\omega^+$. Black profiles denote the smooth-channel flow as a reference.

In Fig. 11, we plot the streamwise mean velocity ($\overline{u}^+$) profiles versus the wall-normal coordinate ($\zeta^+$); note that $\overline{u}$ is averaged over $\xi$, $z$ planes and time, as commonly done in wall-normal oscillation studies (Nabae et al. 2025; Zhang et al. 2024; Nabae et al. 2020). Consistent with the DR data (Fig. 10a,b), there is a negligible change in the mean velocity profile for streamwise waves with respect to the smooth-channel (reference) case (Fig. 11a,b). However, the profiles of the upstream wave (Fig. 11a, inset) predominantly fall below the reference profiles, indicating their drag increase, while the profiles of the downstream wave slightly fall above the reference profile (Fig. 11b, inset), indicating their slight drag reduction. The profiles for the spanwise wave (Fig. 11c) have a significant upward shift, consistent with their significant drag reduction (Fig. 10c,f). Their upward shift is accompanied by thickening of the viscous sublayer, similar to other drag-reducing mechanisms such as spanwise wall oscillation (Quadrio et al. 2009; Rouhi et al. 2025) and riblets (Endrikat et al. 2021b; Rouhi et al. 2022).

## 3.4. Turbulence statistics

We calculate the turbulence statistics following the triple decomposition (9). The turbulence statistics of the upstream and downstream waves are similar, consistent with their marginal changes in drag. Therefore, we only present the turbulence statistics for the upstream waves and for the spanwise waves (Fig. 12 and 13). The r.m.s of $\tilde{u}$ and $\tilde{w}$ profiles (Fig. 12a and 13c) reflect the oscillating wind resulting from the wall deformation for the streamwise and spanwise waves, respectively, as discussed in §3.1. For the streamwise wave, the streamwise wind manifests in a significant $\tilde{u}^+_{rms}$ in the order of $\overline{u}^+$ (Fig. 12a). For the spanwise wave, the wind oscillates in the spanwise direction, resulting in a significant $\tilde{w}^+_{rms}$ (Fig. 13c). The wind profile in the bulk region ($\zeta^+ > 10$) could be derived from the continuity equation for the harmonic field (Touber and Leschziner 2012).

$$\frac{\partial \tilde{u}_i}{\partial x_i} + \frac{\partial \tilde{v}}{\partial y} = 0,$$

where i=1 for the streamwise wave ($\tilde{u}_1 \to \tilde{u}, x_1 \to x$) and i=3 for the spanwise wave ($\tilde{u}_3 \to \tilde{w}, x_3 \to z$). Taking the integral with respect to the wall-normal coordinate yields:

$$\frac{\partial}{\partial x_i} \int_0^h \tilde{u}_i \mathrm{d}y + \tilde{v}_h - \tilde{v}_0 = 0, \tag{11}$$

where $\tilde{v}_h = 0$ and $\tilde{v}_0 = \partial\eta/\partial t$ are the vertical velocities at the bottom and top boundaries, respectively. Fig. 12(a) and 13(c) highlight that $\tilde{u}$ (for upstream wave) and $\tilde{w}$ (for spanwise wave) remain constant beyond $\zeta^+ = 10$. This range accounts for 95% of the entire channel height. Therefore, the integral in (11) can be simplified to

$$\int_0^h \tilde{u}_i dy \approx \tilde{u}_i(x_i, t)h. \tag{12}$$

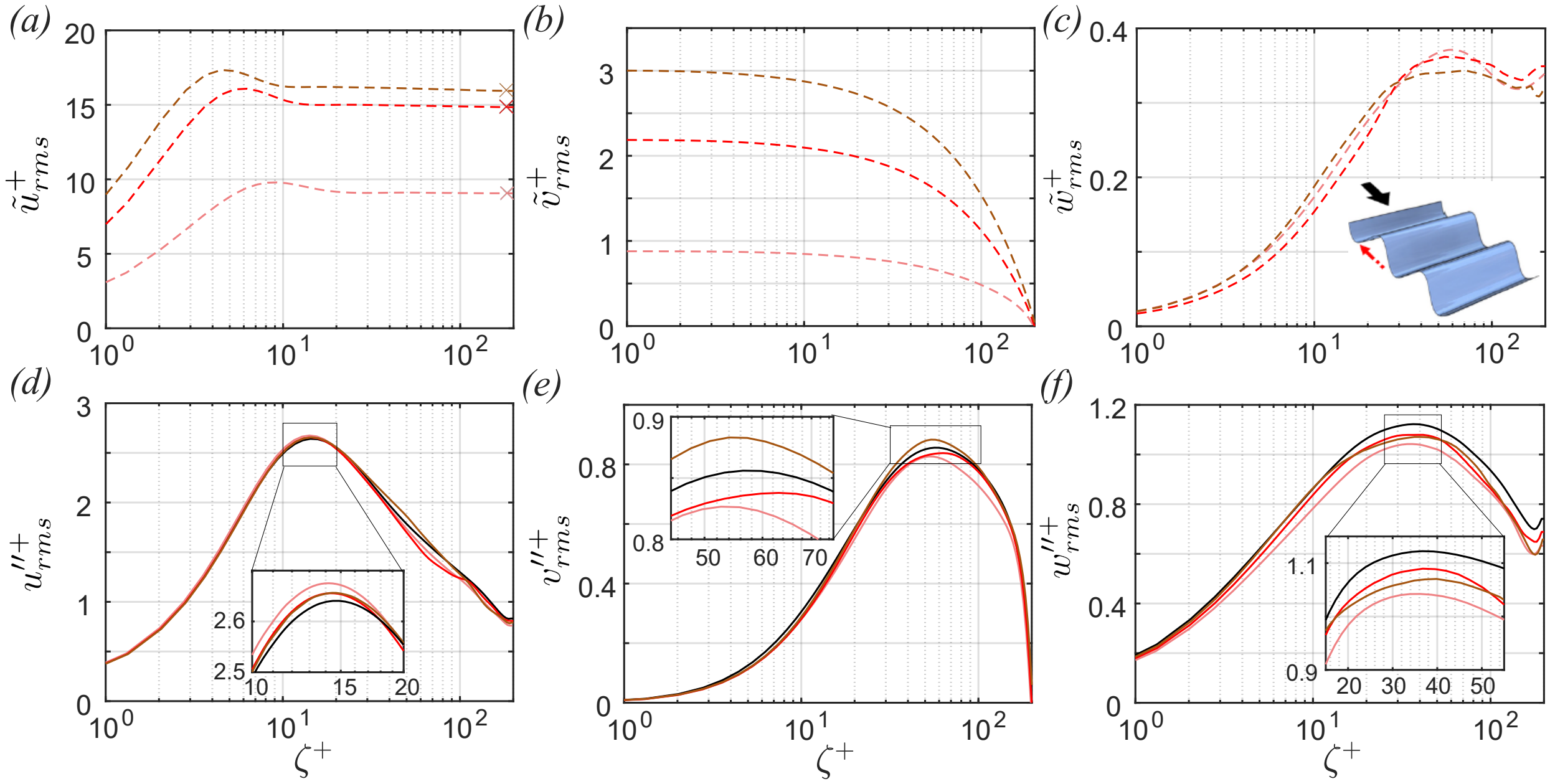

**Figure 12:** Turbulent statistics for upstream waves. (a) Streamwise, (b) wall-normal, and (c) spanwise harmonic Reynolds normal stress. (d) Streamwise, (e) wall-normal, and (f) spanwise turbulent Reynolds normal stress. Profile colors transition from pink to red to brown, representing increasing values of $\eta^+_{\max}\omega^+$. Black profiles denote the smooth-channel flow as a reference. '×' symbols in panel (a) denote the calculation $\tilde{u}^+_{rms}$ based on (15).

Substituting (12) in (11), and after some recasting, we arrive at

$$\frac{\partial \tilde{u}_i}{\partial x_i} = \frac{\tilde{v}_0}{h} = \frac{1}{h}\frac{\partial \eta}{\partial t}. \tag{13}$$

By substituting the expression for $\eta$ from (6), we can obtain the wind velocity in the bulk region

$$\tilde{u}_i = -\frac{\omega}{\kappa h}\left[\eta_1\cos(\kappa x_i - \omega t + \phi_1) + (-1)^n \eta_2 \cos(\kappa x_i - (-1)^n \omega t + \phi_2)\right]. \tag{14}$$

By taking the r.m.s of (14) and viscous scaling, we obtain

$$\tilde{u}^+_{i\ \mathrm{rms}} = \frac{\omega^+}{\kappa^+ Re_\tau}\left[\frac{1}{2}(\eta_1^{+2} + \eta_2^{+2}) + \eta_1^+\eta_2^+\cos(\phi_2 - \phi_1)\right]^{\frac{1}{2}} = \frac{\omega^+\eta^+_{\max}}{\sqrt{2}\kappa^+ Re_\tau}. \tag{15}$$

The semi-analytical relation (15) agrees well with the asymptotic values of the $\tilde{u}^+_{\mathrm{rms}}$ and $\tilde{w}^+_{\mathrm{rms}}$ at the top of the half-channel in Fig. 12(a) and 13(c) (× symbols at $\zeta^+ = 200$). Considering (15), $\tilde{u}^+_{i\ \mathrm{rms}}$ is proportional to $\eta^+_{\max}\omega^+$. For streamwise waves, increasing $\eta^+_{\max}\omega^+$ (pink to red to brown profiles) leads to higher $\tilde{u}^+_{rms}$ (Fig. 12a). Similarly, for spanwise waves, greater $\eta^+_{\max}\omega^+$, results in higher $\tilde{w}_{\mathrm{rms}}$ (Fig. 13c). This outcome explains the clearer trend between DR% and $\eta^+_{\max}\omega^+$ (Fig. 10d,e,f), as DR is related to the oscillating wind strength.

The resulting oscillating wind imposes an oscillating strain in the near-wall region ($1 \le \zeta^+ \le 10$) that interacts with the near-wall turbulence. For the spanwise wave, the imposed strain ($\partial\tilde{w}/\partial y$) is perpendicular to the direction of the near-wall streaks, similar to the Stokes layer strain for the spanwise wall oscillation (Quadrio 2011; Touber and Leschziner 2012; Rouhi et al. 2023). The efficacy of the perpendicular oscillating strain in attenuating the near-wall turbulence has been well reported in the literature (Rouhi et al. 2023). This significant attenuation is also evident in the profiles of $u''^+_{rms}$ for our spanwise wave cases (Fig. 13d), leading to the significant DR (Fig. 10c,f). In the case of the streamwise wave, despite the significant oscillating wind, the $u''^+_{rms}$ profiles do not change compared to the reference smooth case (Fig. 12d). As a result, the streamwise waves marginally change the wall drag (Fig. 10a,b,d,e). Such behaviour is related to the alternating exposure of the near-wall turbulence to local adverse and favourable pressure gradients, as discussed next.

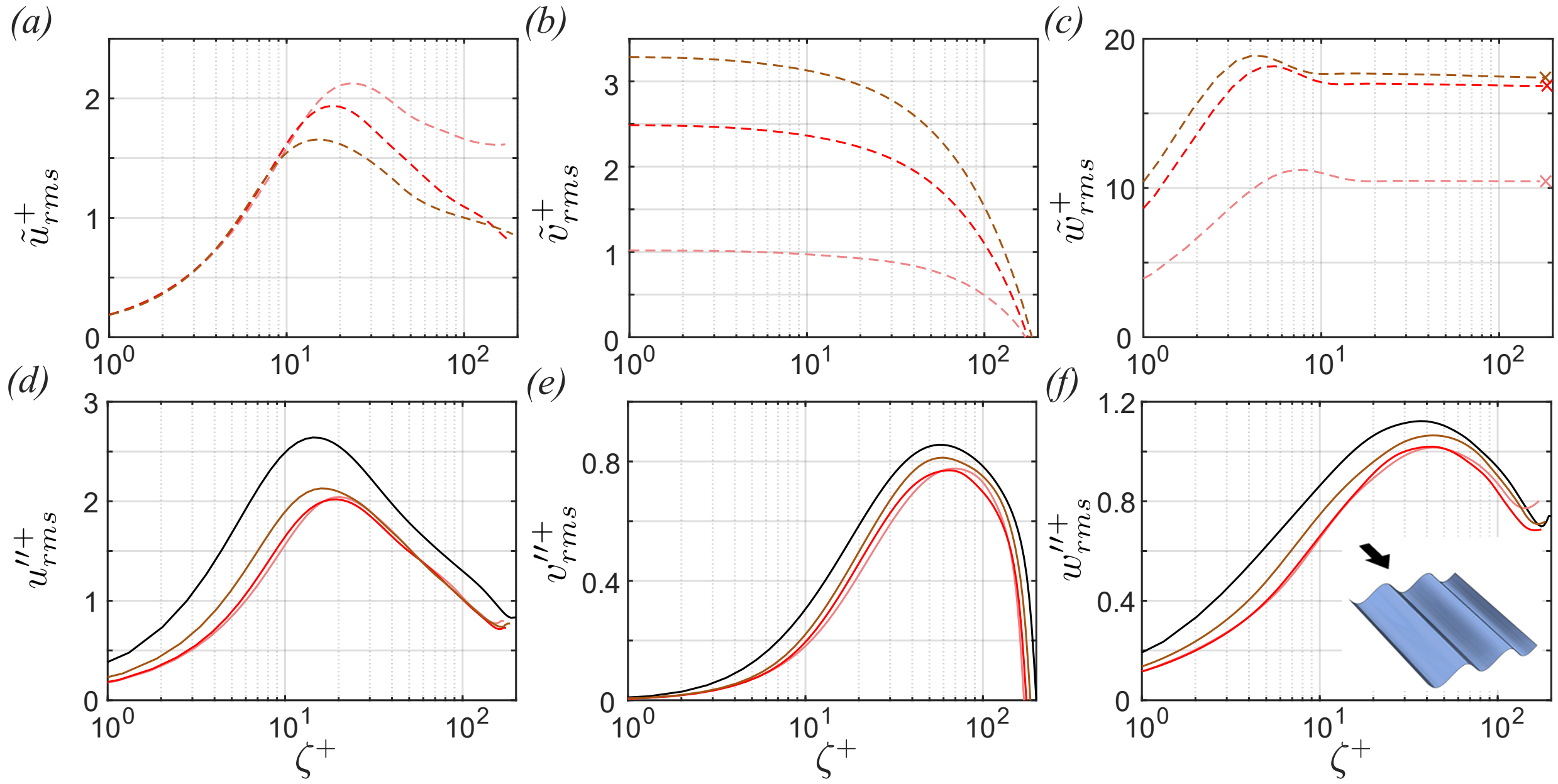

**Figure 13:** Root Mean Square (rms) of Reynolds normal stresses for Spanwise waves. (a) Streamwise, (b) wall-normal, and (c) spanwise harmonic Reynolds normal stress. (d) Streamwise, (e) wall-normal, and (f) spanwise turbulent Reynolds normal stress. Profile colors transition from pink to red to brown, representing increasing values of $\eta^+_{\max}\omega^+$. '×' symbols in panel (c) denote the calculation $\tilde{w}^+_{rms}$ based on (15). Black profiles denote the smooth-channel flow as a reference.

## 3.5. Near-wall turbulence

### *3.5.1. Streamwise waves*

The harmonic pressure gradient due to wall deformation could be deduced from the spanwise-averaged x-momentum equation

$$\frac{\partial\langle u\rangle}{\partial t}+\frac{\partial\langle uu\rangle}{\partial x}+\frac{\partial\langle vu\rangle}{\partial y}=\left(-\frac{1}{\rho}\frac{\partial\langle p\rangle}{\partial x}-\frac{1}{\rho}\frac{dP}{dx}\right)+\nu\left(\frac{\partial^2\langle u\rangle}{\partial x^2}+\frac{\partial^2\langle u\rangle}{\partial y^2}\right). \tag{16}$$

In Fig. 14, we plot the individual terms of (16). In the outer region ($y/h \gtrsim 0.1$), only the transient and pressure gradient terms are dominant (Fig. 14a,d), and (16) will be simplified to

$$\frac{\partial\langle u\rangle}{\partial t}=-\frac{1}{\rho}\frac{\partial\langle p\rangle}{\partial x}-\frac{1}{\rho}\frac{dP}{dx}. \tag{17}$$

Subtracting the $x$-averaged of (17) ($\partial\bar{u}/\partial t=-\partial\bar{p}/\rho\partial x-dP/\rho dx$) from itself, yields

$$\frac{\partial\tilde{u}}{\partial t}=-\frac{1}{\rho}\frac{\partial\tilde{p}}{\partial x}. \tag{18}$$

Substituting for $\tilde{u}$ (14) in (18) and non-dimensionalization, we can obtain

$$\frac{d\tilde{p}^+}{dx^+}=\frac{-\omega^{+2}}{Re_\tau\kappa^+}\left[\eta_1^+\sin(\kappa^+x^+-\omega^+t^++\phi_1)+\eta_2^+\sin(\kappa^+x^+-(-1)^n\omega^+t^++\phi_2)\right]. \tag{19}$$

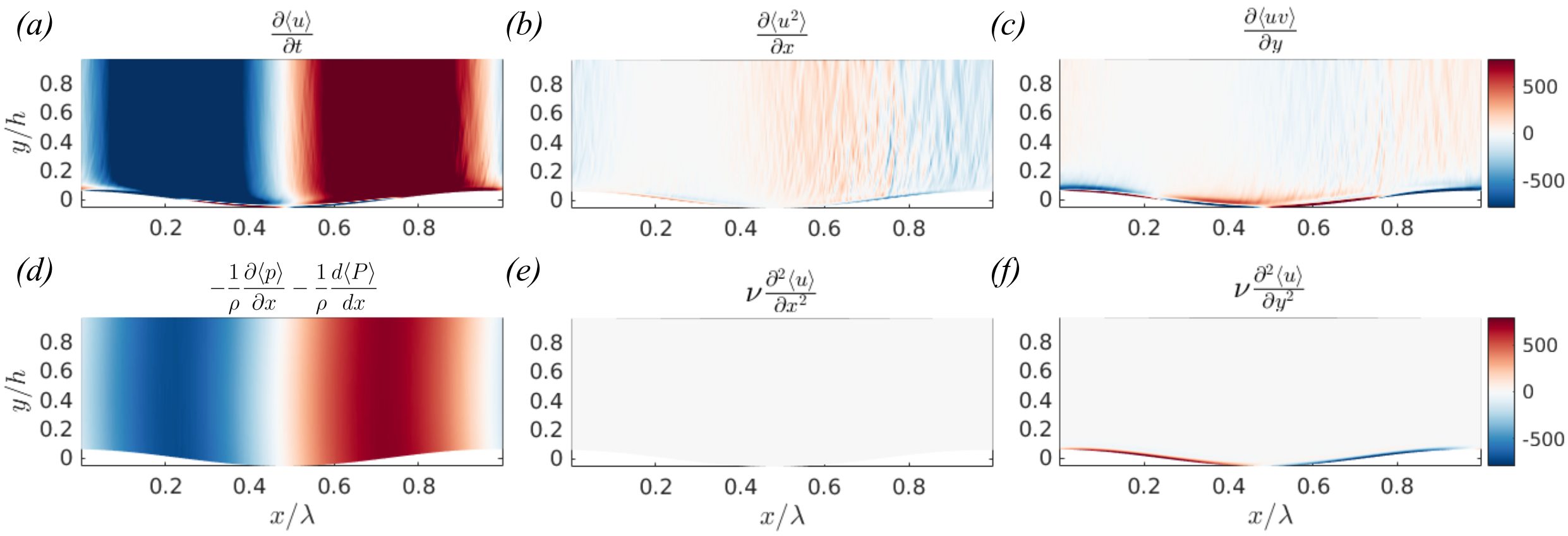

**Figure 14:** Spanwise averaged $x$-momentum budgets for USHW-91 case. (a) Unsteady term; (b) Streamwise convection term; (c) Wall-normal convection term; (d) Pressure terms; (e) Streamwise diffusion term; (f) Wall-normal diffusion term.

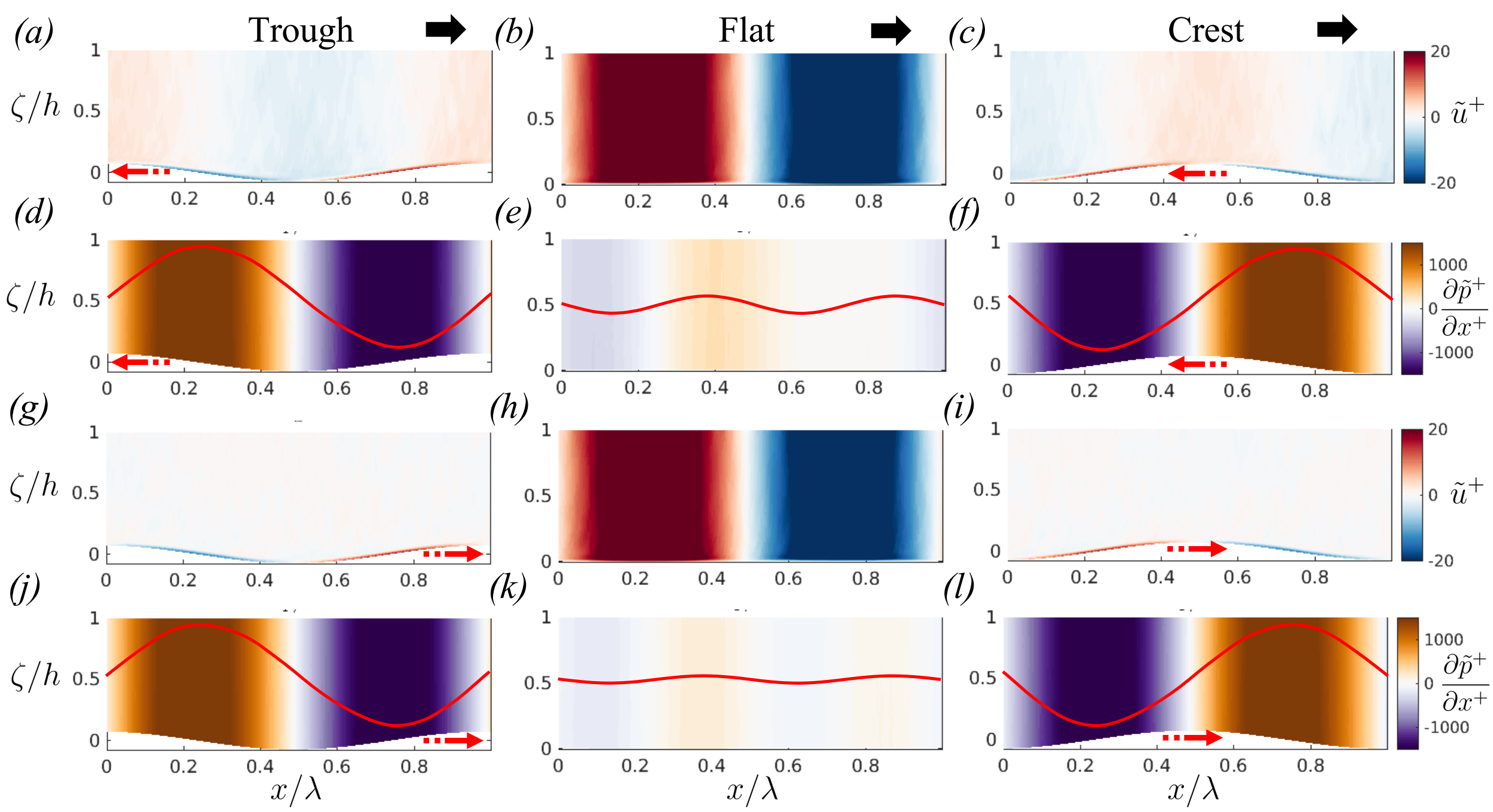

**Figure 15:** Side view contour plots of harmonic streamwise velocity (panels a-c, g-i) and pressure gradient (panels d-f, j-l) over upstream wave (panels a-f) and downstream wave (panels g-l) at wave trough (panels a,d,g,j), flat (panels b,e,h,k), and crest (panels c,f,i,l) configuration. The red profiles in panels (d-f,j-l) present the pressure gradient corresponding to (19).

In Fig. 15, we overlay (19) onto the $d\tilde{p}^+/dx^+$ field from DNS at different phases during oscillation, which agree well with each other. In Fig. 15, we also plot $\tilde{u}$ fields, at the same phases as the $d\tilde{p}^+/dx^+$ fields. For both the upstream wave (Fig. 15 a-f), and the downstream wave (Fig. 15g-l), the near-wall $\tilde{u}$ becomes negative during APG ($d\tilde{p}^+/dx^+ > 0$), indicating the near-wall flow deceleration. On the other hand, during FPG ($d\tilde{p}^+/dx^+ < 0$), the near-wall $\tilde{u}$ becomes positive, indicating the near-wall flow acceleration.

To better study the effect of local APG and FPG on the near-wall turbulence, we conditionally average the spanwise-averaged velocity ($\langle u\rangle^+$), and streamwise turbulent stress ($\overline{u''^{2+}}$) based on FPG ($d\tilde{p}^+/dx^+ < 0$) and APG ($d\tilde{p}^+/dx^+ > 0$) for the upstream and downstream waves (Fig. 16). For the upstream waves, the near-wall flow is reversed due to APG (Fig. 16a, gray area). We also observe a mild flow reversal for the downstream wave during APG (Fig. 16b), but it is significantly milder compared to the upstream wave. This indicates that APG has a stronger influence on the upstream wave than on the downstream counterpart. Furthermore, the conditionally averaged $\overline{u''^{2+}}$ highlights that APG and FPG have a stronger influence on the near-wall turbulence for the downstream wave (Fig. 16d) compared to the upstream one (Fig. 16c). For the upstream wave, the inner peak of $\overline{u''^{2+}}$ is at $\zeta+ = 14$ during both APG and FPG, and is close to the locale of the inner peak of $\overline{u''^{2+}}$ for the reference smooth case. However, for the downstream wave, the near-wall turbulence is more influenced by the local APG and FPG. During FPG, the inner peak of $\overline{u''^{2+}}$ is shifted away from the wall compared to the reference smooth case. The shift in the inner peak of $\overline{u''^{2+}}$ is an indication of drag reduction (Rouhi et al. 2025). On the other hand, during APG, the inner peak of $\overline{u''^{2+}}$ shifts closer to the wall, indicating a local increase in drag. The streamwise wave generates an alternating APG and FPG.

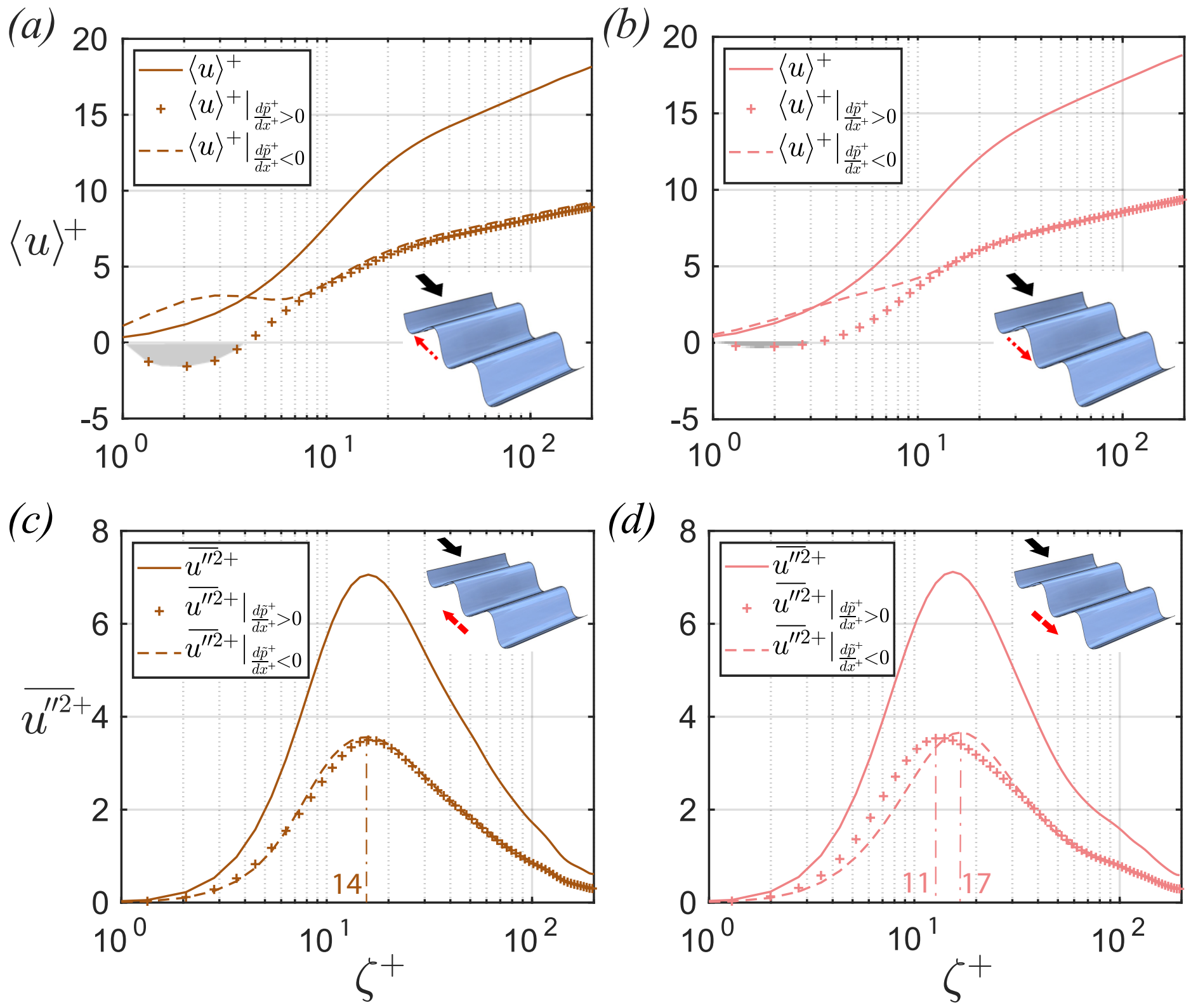

**Figure 16:** Conditional averaged streamwise velocity profile based on adverse and favorable pressure gradient for (a) Upstream wave, (b) Downstream wave. Gray-shaded areas denote negative spanwise-averaged velocities. Conditional averaged streamwise Reynolds normal stress based on adverse and favorable pressure gradient for (c) upstream wave, (d) downstream wave.

The different responses of the near-wall turbulence to the pressure gradient for the upstream and downstream waves could also be interpreted from the near-wall visualizations of the turbulent fluctuations (Fig. 17). Here, we focus on the $u''v''^+$ field, due to its direct connection to the wall drag (Rouhi et al. 2025). The wave-shielded region generally corresponds to the APG, while the wave-exposed region is associated with FPG. For the upstream wave, the $u''v''^+$ respond similarly to APG and FPG (Fig. 17a,b,c), consistent with the similar conditionally averaged $\overline{u''^{2+}}$ profiles (Fig. 16c). In contrast, for the downstream-wave case, the $u''v''^+$ field highlights more energetic and elongated structures during APG, while during FPG the structures are broken into smaller scales and $u''v''^+$ energy is attenuated.

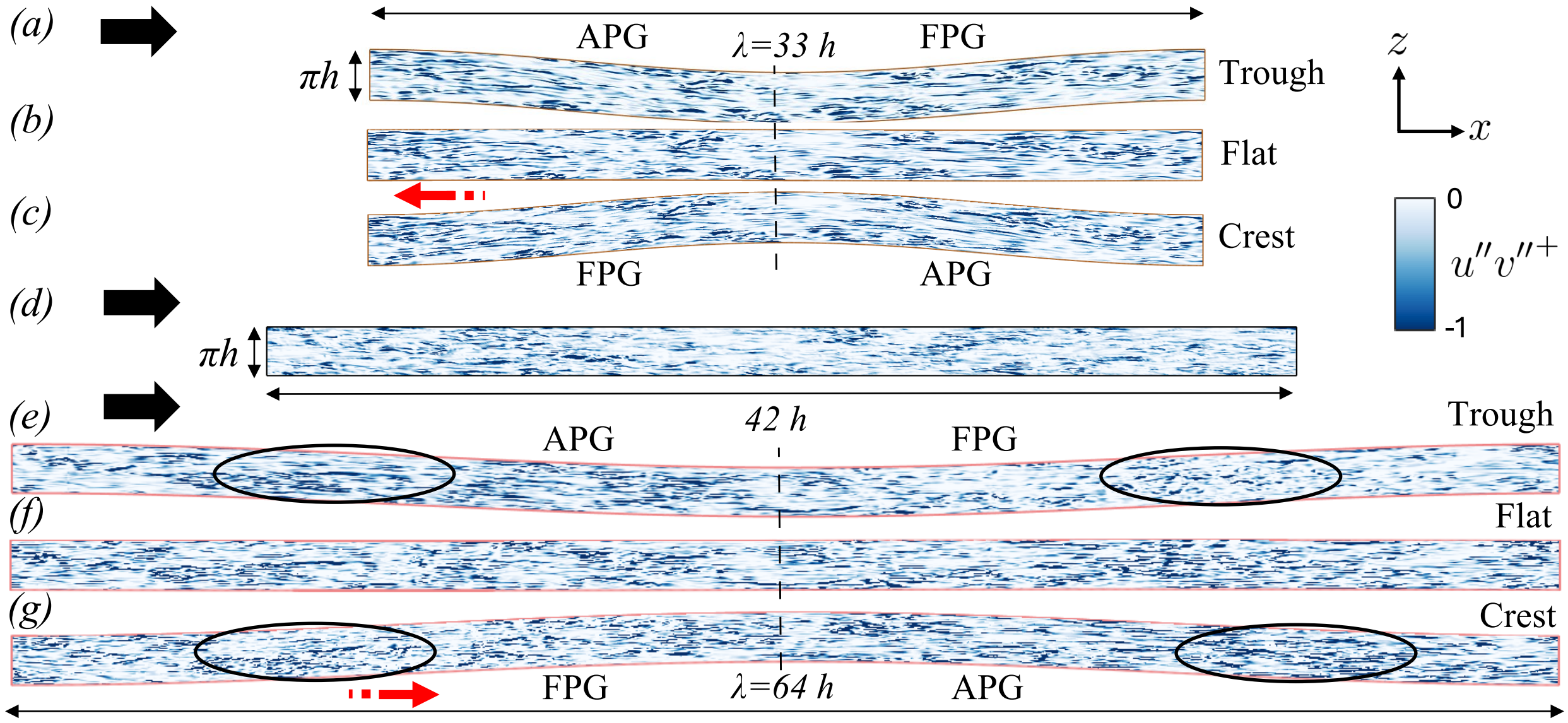

**Figure 17:** Contour plots of Reynolds shear stress at $\zeta^+ = 8$. Panels (a–c) show the upstream wave for the trough, flat, and crest configurations, respectively, while panel (d) presents the smooth-wall reference case. Panels (e–g) display the downstream wave for the trough, flat, and crest configurations, respectively. Regions of APG (wave-shielded) and FPG (wave-exposed) are indicated over the waves.

### *3.5.2. Spanwise waves*

The spanwise waves generate a harmonic spanwise pressure gradient that induces a spanwise wind ($\tilde{w}^+$); the wind induces a transverse alternating shear near the wall, as is evident during the trough and crest phases of the wall deformation (Fig. 18a,c). Additionally, the wall deformation induces low- and high-streamwise-momentum regions (Fig. 18d,e,f).

Fig. 19 compares the near-wall structures in the spanwise-wave cases with those over a smooth wall. Compared to the reference smooth case (Fig. 19a), during the crest and trough configurations, the near-wall turbulence is significantly attenuated (Fig. 19b,d); the level of attenuation is more prominent in the high-momentum regions. On the other hand, during the flat configuration, the near-wall turbulence is energized (Fig. 19c). We observe the emergence of energetic streamwise-elongated structures. Such elongated structures are commonly associated with reduced bursting activity and suppression of the near-wall turbulence regeneration cycle. This behavior indicates that spanwise forcing enhances the spatial regularity and streamwise coherence of the near-wall structures (Shao et al., 2025; Tomiyama and Fukagata, 2013; Bai et al., 2014). Overall, the attenuation and elongation of the near-wall structures suggest weakened near-wall turbulent activity and reduced Reynolds shear stress, thereby contributing to drag reduction.

## 4. Conclusions

Direct numerical simulations of turbulent half-channel flow were performed at friction Reynolds number of $Re_\tau = 200$. We investigated the impact of actively generated wall deformations, produced via piezoelectric actuation, on turbulent skin-friction drag. We explored actuation across practical piezoelectric operating ranges (voltage range $250 \sim 500V$ and frequency range $119 \sim 543Hz$). By assuming that the piezoelectric setup is installed in the experimental configuration of Ding et al. (2024), the actuation parameters in viscous units are $2 \leq \eta^+_{\max} \leq 34$ and $-0.58 \leq \omega^+ \leq 0.70$. Unidirectional travelling, standing, and hybrid waves were generated via wall deformations. The waves were imposed in the upstream, downstream, and spanwise directions as bottom-wall boundary conditions of the turbulent channel flow. Although the upstream and downstream waves produced mild drag reductions of $-4\% \sim 6\%$, the hybrid spanwise wave achieved up to 27% drag reduction at $\omega^+ = 0.17$ and $\eta^+_{\max} = 11.5$. The mean velocity profiles for the streamwise waves exhibited only minor shifts, whereas the spanwise waves produced an upward shift of the mean profile and a thickening of the viscous sublayer, similar to other near-wall flow controlling mechanisms (riblets, spanwise plane oscillation). Turbulence statistics were weakly affected by the streamwise waves, but they were significantly attenuated by the spanwise waves.

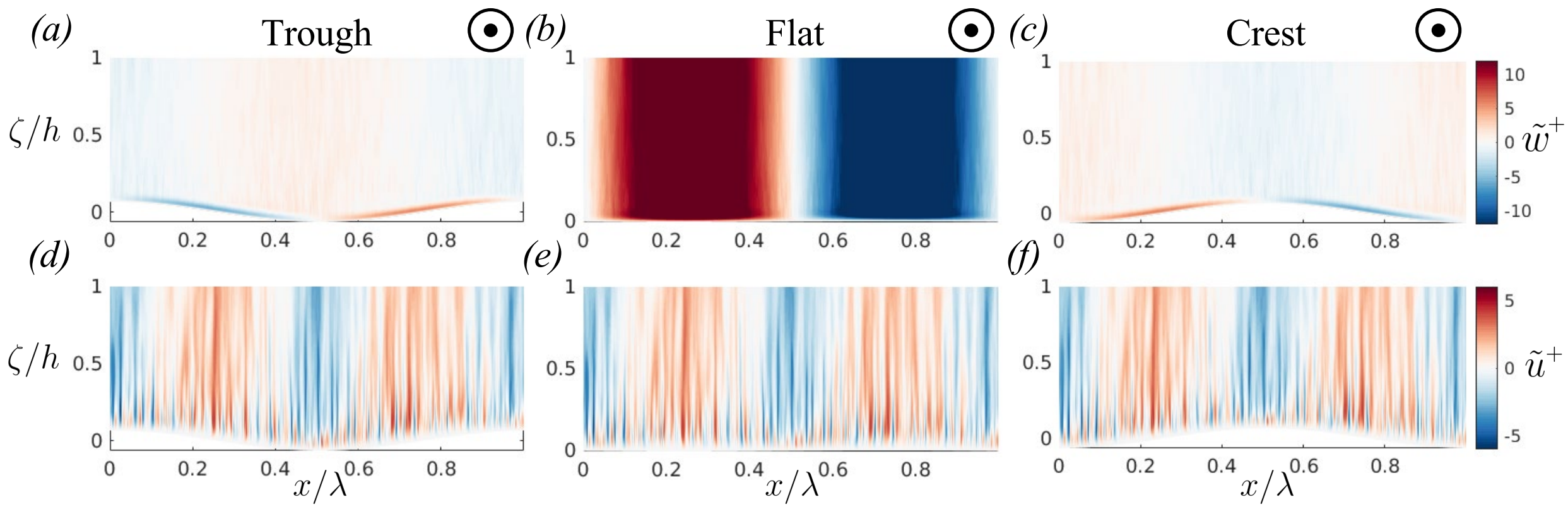

**Figure 18:** Front view contour plots of harmonic spanwise velocity (panels a,b,c) and streamwise velocity (panels d,e,f) over spanwise wave at trough (panels a,d), flat (panels b,e), and crest (panels c,f) configuration.

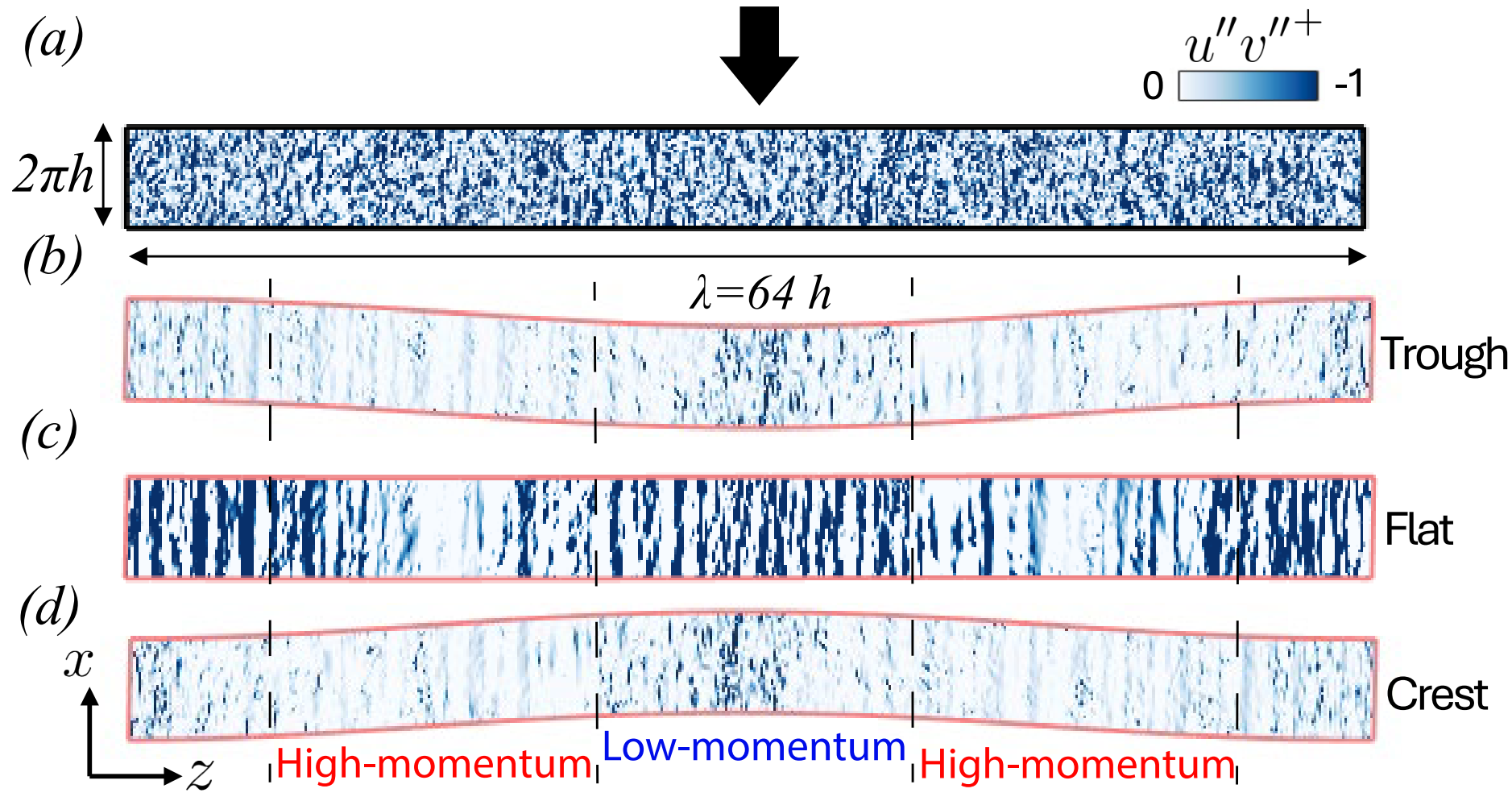

**Figure 19:** Contour plots of Reynolds shear stress for (a) the smooth-wall reference case, (b) the spanwise wave at the wave trough, (c) the spanwise wave in its locally flat configuration, and (d) the spanwise wave at the wave crest, all plotted at $\zeta^+ = 15$.

Analysis of the harmonic velocity and pressure revealed that wall deformation generates an oscillating wind, as well as an oscillating pressure gradient. For the streamwise wave, the oscillating pressure gradient consists of an alternating adverse pressure gradient (APG, $\partial\tilde{p}/\partial x > 0$), and favourable pressure gradients (FPG, $\partial\tilde{p}/\partial x < 0$). Conditional averaging based on APG and FPG, highlighted that drag is locally increased and decreases during APG and FPG, respectively. As a result, the total drag change over the streamwise waves is marginal.

For the spanwise waves, the spanwise wind generated a transverse alternating shear layer, which in turn produced distinct high- and low-streamwise-momentum regions. Depending on the oscillation phase, the near-wall turbulence was either attenuated or energized with the emergence of coherent streamwise-elongated structure. Overall, owing to the dominance of the near-wall turbulence attenuation, significant drag reduction (up to 27%) was achieved. Our present study highlights the potential of piezoelectric actuation for significant drag reduction by generating multi-mode surface deformations.

## Acknowledgments

We thank EPSRC for the computational time made available on ARCHER2 via the UK Turbulence Consortium (EP/X035484/1), and the UKRI access to the HPC call 2024.

## CRediT authorship contribution statement

**Amir Amjadimanesh:** Conceptualization, Methodology, Software, Validation, Formal analysis, Investigation, Visualization, Writing - original draft.
**Aman Kidanemariam:** Methodology, Writing - review & editing, Supervision.
**David Chappell:** Methodology, Writing - review & editing, Supervision.
**Mahdi Bodaghi:** Conceptualization, Writing - review & editing, Supervision.
**Amirreza Rouhi:** Conceptualization, Methodology, Writing - review & editing, Supervision, Project administration.

## Declaration of competing interest

The authors declare that they have no known competing financial interests or personal relationships that could have appeared to influence the work reported in this paper.

## Data availability

The data that support the findings of this study are available from the corresponding author upon reasonable request.

## A. Validations

The validation of computational fluid dynamics (CFD) codes is an essential process to ensure the accuracy and reliability of simulations. This section details the systematic validation of the in-house Direct Numerical Simulation (DNS) code, MELA, used in this research. Validation is performed in three progressive stages: I. Simple channel turbulent flow, II. channel flow over a bottom wavy wall (standing wave), and III. Channel flow over a travelling wave bottom wall (§2.2).

In the first stage, the simple channel turbulent flow is chosen due to its well-documented characteristics and availability of benchmark data. The simulation setup involves a straight channel with periodic boundary conditions in the streamwise and spanwise directions, and no-slip conditions at the walls. The validation focuses on comparing the velocity profiles and Reynolds normal stress distributions with those reported in the literature from DNS. Achieving agreement with these benchmarks confirms the code's capability to accurately resolve canonical turbulent flows.

The second stage extends the validation to a more complex geometry by introducing a bottom wavy wall with a standing wave pattern. This scenario examines the code's ability to handle complex boundary conditions and flow interactions. The sinusoidal wavy bottom wall introduces additional flow dynamics, and the validation involves comparing the mean velocity profile and Reynolds normal stresses against numerical data from previous studies.

In the final stage, validation proceeds to a channel with a bottom wall described by a travelling wave. This introduces time-dependent boundary conditions and tests the code's robustness in capturing unsteady flow phenomena. The validation criteria include comparisons of mean velocity and analyses of turbulent structures and their interactions with the travelling wave. The results of this stage were reported in §2.2.

Each validation stage builds on the previous one, incrementally increasing the complexity of the flow scenarios to thoroughly test the DNS code in MELA. Successful validation across these stages demonstrates the code's reliability and accuracy, providing a solid foundation for its application to more complex and novel flow problems in the subsequent sections of this research.

### A.1. Smooth channel flow

We well-validated our solver for smooth channel flow (Moser et al. 1999). Fig. 20 shows the comparison of our half-channel flow statistics with Moser et al. (1999)'s full-channel at $Re_\tau = 395$. The domain size is $L_x \times L_y \times L_z = 2\pi h \times h \times \pi h$. We performed a systematic grid-convergence study by running calculations at multiple grid resolutions (Table 3).

| Grid title | Grid number $(N_x \times N_y \times N_z)$ | $Re_\tau$ | $\Delta_x^+$ | $\Delta_y^+$ | $\Delta_z^+$ | $C_f$ | Error Percentage |
|---|---|---|---|---|---|---|---|
| Moser's DNS | $256 \times 193 \times 192$ | 395 | 9.8 | [..., 6.5] | 6.5 | 0.00658 | - |
| Coarse | $192 \times 96 \times 192$ | 395 | 13.1 | [0.6,8.7] | 6.5 | 0.00662 | 0.60% |
| Fine | $256 \times 128 \times 256$ | 395 | 9.8 | [0.4,6.5] | 4.9 | 0.00660 | 0.30% |

**Table 3**
Quantitative validation of skin friction coefficient for different cases. The runtime for all of our cases is 40 large-eddy turnover times. The lower limit of Moser's $\Delta_y^+$ has not been reported.

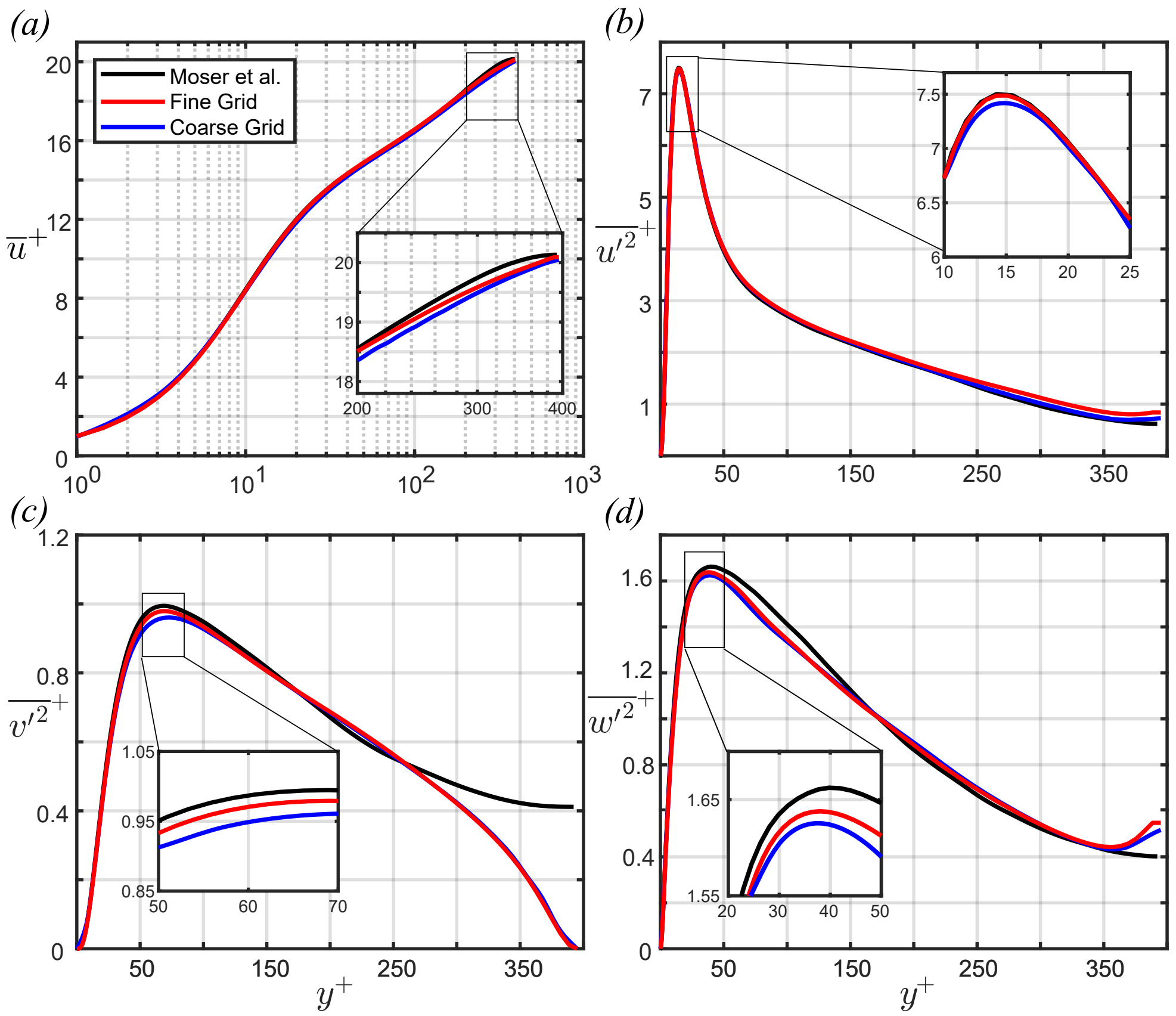

**Figure 20:** Validation of DNS solver for half-channel flow at $Re_\tau = 395$. (a) streamwise mean velocity profile (b) streamwise fluctuating stress (c) wall-normal fluctuating stress (d) spanwise fluctuating stress.

We compare our results with the DNS of Moser et al. (1999) for turbulent channel flow at matched $Re_\tau = 395$ (Fig. 20). We obtain an excellent agreement in the mean velocity and the streamwise fluctuating stress profiles (Fig. 20a,b). We observe that the mean velocity profiles for both the coarse and fine grids agree quite well with the reference DNS. Similarly, both grids agree quite well with the reference DNS in the fluctuating stress profiles, especially in capturing the inner peak. Fig. 20(c) checks the validity of the fluctuating stress profile in the wall-normal direction. Our simulation is in good agreement with Moser's data except in the tail of the curve. This slight difference is due to the change in the top boundary conditions. We applied the symmetry boundary condition at the top boundary, while Moser et al. solved the full channel flow. Fig. 20(d) demonstrates the fluctuating stress profile in the spanwise direction. We further quantified the accuracy of our calculation by comparing the skin-friction coefficient $C_f$ between our cases and those of Moser et al. (1999) (Table 3). The $C_f$ value for the fine grid differs by less than 1% from that reported by Moser et al. (1999). The results are in very good agreement with them, even with a coarse grid, except in the outer region, where differences between the half-channel and full-channel boundary conditions are responsible.

| Grid title | Grid number ($N_x \times N_y \times N_z$) | $\Delta_x^+$ | $\Delta_y^+$ | $\Delta_z^+$ | $u_\tau$ Top | $u_\tau$ Bottom | $C_f$ Bottom | Error |
|---|---|---|---|---|---|---|---|---|
| Maass's DNS | $256 \times 96 \times 128$ | 10.9 | [1.6,...] | 10.9 | 0.070 | 0.104 | 0.0216 | - |
| Coarse | $96 \times 96 \times 96$ | 30.1 | [1.5,16.3] | 15.1 | 0.068 | 0.107 | 0.0228 | 5.5% |
| Fine | $128 \times 128 \times 128$ | 22.2 | [1.1,10.6] | 11.1 | 0.069 | 0.105 | 0.0220 | 1.8% |

**Table 4**
Quantitative validation of the friction coefficient for different cases. The runtime for all of our cases is 60 large-eddy turnover times. The upper limit of Maass's $\Delta_y^+$ has not been reported.

### A.2. Channel flow with bottom wavy wall

As the second benchmark, we validated the solver with the static wavy wall study of Maaß and Schumann (1996) in curvilinear coordinates $(\xi, \zeta, z)$ (Fig. 21). They considered channel flow with a bottom static wavy wall at $Re_b = 6760$. The domain size is $L_x \times L_z \times h = 4 \times 2 \times 1$. We conducted a systematic grid-convergence study across different resolutions. We obtain excellent agreement in the mean velocity and the streamwise fluctuating stress profiles (Fig. 21a,b). We observe that, for the coarse-grid case, the mean velocity profiles agree quite well with the reference DNS. Similarly, the fluctuating stress profiles agree quite well with the reference DNS, especially in capturing the inner peak. Fig. 21(c) checks the validity of the fluctuating stress profile in the wall-normal direction. Fig. 21(d) demonstrates the fluctuating stress profile in the spanwise direction.

We further quantified the accuracy of our calculation by comparing the friction coefficient $C_f$ in our cases with that reported in Maaß and Schumann (1996). Table 4 presents the grid resolutions and the related friction coefficients. Improving the grid resolution reduces the error percentage, as observed in Maaß and Schumann (1996)'s investigation. For the fine grid, the error is less than 2%, confirming a good agreement of friction coefficients with Maaß and Schumann (1996)'s data.

Therefore, we assure the validity and accuracy of our numerical approach. Given that the phase speed of piezoelectric actuation falls within the range of slow and intermediate travelling waves, we can confidently assert that our DNS setup is reliable for this research.

## B. Simulation cases

The simulation cases are summarized in the Table 5.

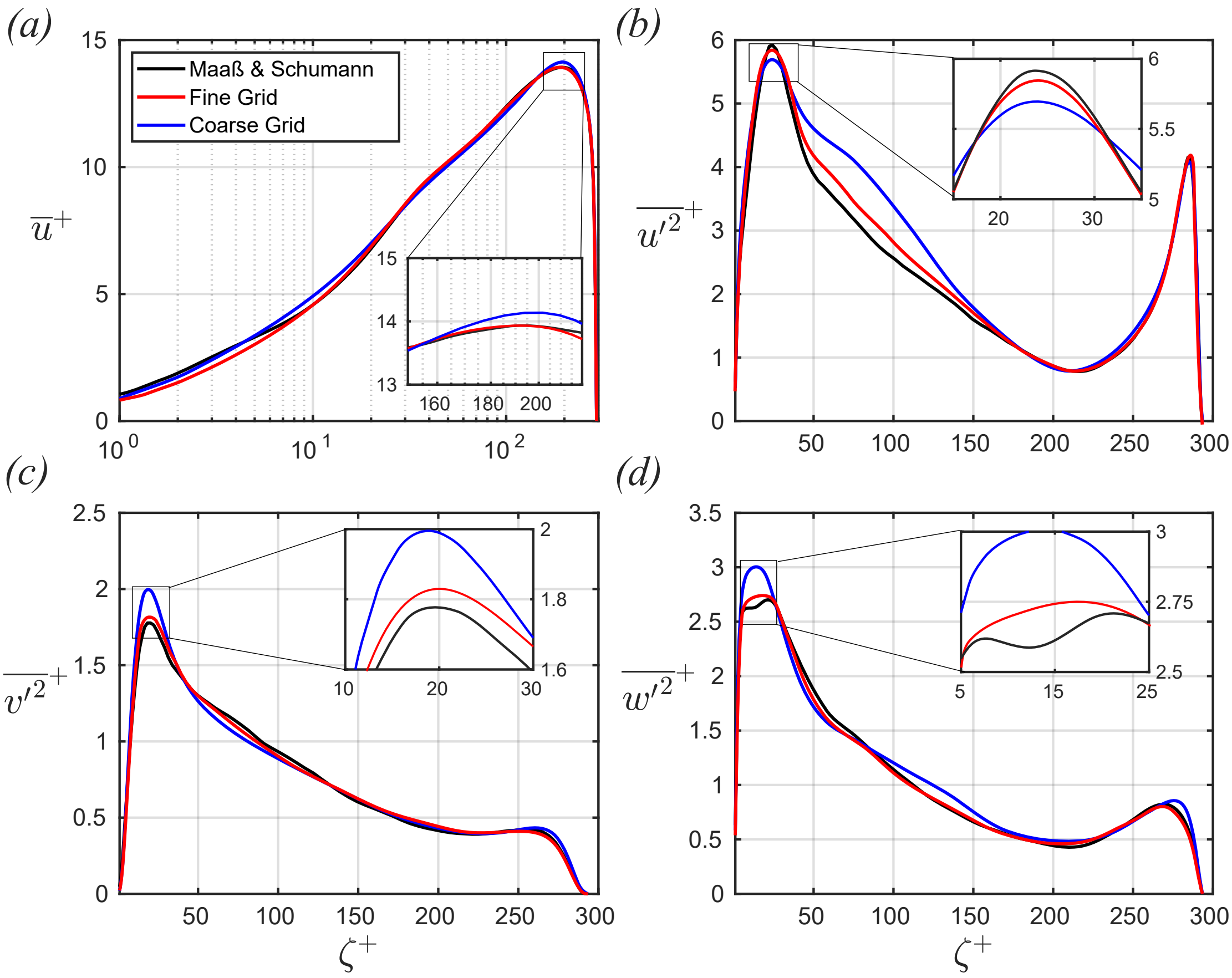

**Figure 21:** Validation of DNS solver for full-channel flow with bottom wavy wall at $Re_\tau = 215$. (a) streamwise mean velocity profile (b) streamwise fluctuating stress (c) wall-normal fluctuating stress (d) spanwise fluctuating stress.

| $Wave$ ($CF$[%]) | $\kappa^+ \times 10^4$ | $\omega^+$ | $\eta_{1\,max}^+, \eta_{2\,max}^+$ | $\phi_1, \phi_2$ | $Re_\tau$ | $L_x/h$ | $\Delta x^+, \Delta z^+$ | $DR$[%] |
|---|---|---|---|---|---|---|---|---|
| DSHW-90* | 5.01 | 0.14 | 6.81,6.15 | 0.73,-0.83 | 196 | 63.79 | (8.3,4.9) | 3.2 |
| DSHW-86 | 7.25 | 0.28 | 4.82,4.17 | 0.66,-0.90 | 196 | 44.22 | (8.6,4.9) | 5.3 |
| DSTW-0.5 | 7.18 | 0.28 | 1.45,0.78 | 0.51,-1.05 | 198 | 44.46 | (8.7,4.9) | 2.1 |
| DSTW-2.5 | 7.61 | 0.31 | 3.36,2.70 | -0.70,0.87 | 194 | 44.78 | (8.7,4.9) | 4.3 |
| DSHW-91** | 7.58 | 0.31 | 7.50,6.87 | -0.70,0.87 | 196 | 42.32 | (8.3,4.9) | 3.4 |
| DSHW-93 | 7.62 | 0.32 | 10.09,9.46 | -0.78,0.78 | 194 | 42.40 | (8.3,4.9) | 5.4 |
| DSSW-95 | 7.47 | 0.30 | 12.47,11.84 | -0.78,0.78 | 198 | 42.59 | (8.3,4.9) | 2 |
| DSSW-96 | 7.34 | 0.29 | 16.57,15.92 | -0.78,0.78 | 202 | 42.59 | (8.3,4.9) | -1.5 |
| DSHW-92.5 | 8.70 | 0.41 | 7.79,7.18 | 0.71,-0.86 | 194 | 37.16 | (9.7,4.9) | 5.4 |
| DSHW-92*** | 9.61 | 0.50 | 6.31,5.78 | -0.78,0.78 | 197 | 33.23 | (8.7,4.9) | 2.7 |
| DSHW-94 | 10.73 | 0.63 | 6.96,6.59 | 0.77,-0.79 | 196 | 29.92 | (7.8,4.9) | 4.3 |
| USHW-90* | 4.95 | −0.13 | 6.89,6.22 | 0.73,-0.83 | 199 | 63.79 | (8.3,4.9) | 1.6 |
| USHW-86 | 7.05 | −0.27 | 4.87,4.22 | 0.66,-0.90 | 200 | 44.22 | (8.6,4.9) | 0.4 |
| USTW-0.5 | 7.25 | −0.29 | 1.43,0.77 | 0.51,-1.05 | 196 | 44.46 | (8.7,4.9) | 0.5 |
| USTW-2.5 | 7.41 | −0.30 | 3.42,2.77 | -0.70,0.87 | 199 | 44.78 | (8.7,4.9) | 1.2 |
| USHW-91** | 7.36 | −0.29 | 8.59,7.86 | -0.70,0.87 | 201 | 42.32 | (8.3,4.9) | 0.1 |
| USHW-93 | 7.27 | −0.29 | 10.59,9.93 | -0.78,0.78 | 203 | 42.40 | (8.3,4.9) | -1.8 |
| USSW-95 | 7.23 | −0.28 | 13.16,12.49 | -0.78,0.78 | 205 | 42.59 | (8.3,4.9) | -2.7 |
| USSW-96 | 7.25 | −0.29 | 17.31,16.64 | -0.78,0.78 | 204 | 42.59 | (8.3,4.9) | -1.3 |
| USHW-92.5 | 8.32 | −0.38 | 8.12,7.48 | 0.71,-0.86 | 204 | 37.16 | (9.7,4.9) | -2.4 |
| USHW-92*** | 9.30 | −0.47 | 6.69,6.13 | -0.78,0.78 | 204 | 33.23 | (8.7,4.9) | -2.7 |
| USHW-94 | 10.29 | −0.56 | 7.38,6.99 | 0.77,-0.79 | 204 | 29.92 | (7.8,4.9) | -3 |
| $Wave$ ($CF$[%]) | $\kappa^+ \times 10^4$ | $\omega^+$ | $\eta_{1\,max}^+, \eta_{2\,max}^+$ | $\phi_1, \phi_2$ | $Re_\tau$ | $L_z/h$ | $\Delta x^+, \Delta z^+$ | $DR$[%] |
| SPHW-90* | 5.80 | 0.18 | 6.02,5.43 | 0.73,-0.83 | 170 | 63.79 | (9.8,6.2) | 27.6 |
| SPHW-86 | 7.69 | 0.32 | 4.54,3.93 | 0.66,-0.90 | 185 | 44.22 | (9.8,4.3) | 15 |
| SPTW-0.5 | 7.40 | +0.29 | 1.42,0.76 | 0.51,-1.05 | 194 | 44.46 | (9.8,4.3) | 6.6 |
| SPTW-2.5 | 7.82 | +0.33 | 3.23,2.62 | -0.70,0.87 | 189 | 44.78 | (9.8,4.2) | 11.3 |
| SPHW-91 | 8.40 | 0.38 | 6.90,6.30 | -0.70,0.87 | 177 | 42.32 | (9.8,4.1) | 22.2 |
| SPHW-93** | 8.34 | 0.38 | 9.20,8.64 | -0.78,0.78 | 178 | 42.40 | (9.8,4.1) | 21.3 |
| SPSW-95 | 8.32 | 0.38 | 11.27,10.69 | -0.785,0.78 | 177 | 42.59 | (9.8,4.2) | 21.7 |
| SPDW-96 | 8.60 | 0.40 | 15.04,14.45 | -0.78,0.785 | 172 | 42.59 | (9.8,4.2) | 22.6 |
| SPHW-92.5 | 9.24 | 0.47 | 7.31,6.75 | 0.71,-0.86 | 183 | 37.16 | (9.8,4.8) | 16.7 |
| SPHW-92*** | 10.30 | 0.57 | 6.03,5.52 | -0.78,0.78 | 184 | 33.23 | (9.8,4.3) | 15.4 |
| SPHW-94 | 11.25 | 0.69 | 6.63,6.27 | 0.77,-0.79 | 186 | 29.92 | (9.8,3.9) | 13.3 |

**Table 5**

Summary of parameters for production runs. $L_z/h = \pi$, for the streamwise waves, while $L_x/h = 2\pi$ for spanwise waves. Cases marked with asterisks (*) are delegate cases. The symbols *, **, and *** represent the pink, red, and brown profiles, respectively.